\documentclass[twocolumn]{aastex7}

\usepackage{amsmath}
\usepackage{enumitem}
\usepackage{graphicx}
\usepackage{subcaption}
\newcommand{\Msun}{{\rm \,M_\odot}}

\newcommand{\Rsun}{{\rm \,R_\odot}}
\newcommand{\Zsun}{{\rm \,Z_\odot}}
\newcommand{\new}[1]{\textcolor{black}{#1}}

\begin{document}

\title{\new{Orbital Eccentricity And Spin-Orbit Misalignment} Are Evidence That\\Neutron Star-Black Hole Mergers Form Through Triple Star Evolution}

\author{Jakob Stegmann}
\altaffiliation{Both authors contributed equally.\\Email: jstegmann@mpa-garching.mpg.de; jklencki@mpa-garching.mpg.de}
\affiliation{Max Planck Institute for Astrophysics, Karl-Schwarzschild-Straße 1, 85748 Garching, Germany}
\email[show]{}  

\author{Jakub Klencki} 
\altaffiliation{Both authors contributed equally.\\Email: jstegmann@mpa-garching.mpg.de; jklencki@mpa-garching.mpg.de}
\affiliation{Max Planck Institute for Astrophysics, Karl-Schwarzschild-Straße 1, 85748 Garching, Germany}
\email[show]{}

\begin{abstract}
There is growing evidence that a substantial fraction of the neutron star-black holes (NSBHs) detected through gravitational waves merge with non-zero eccentricity or large BH spin-orbit misalignment. This is in tension with leading formation scenarios to date. Residual eccentricity rules out formation from isolated binary stars, while NS natal kicks and unequal masses of NSBHs inhibit efficient pairing in dense stellar environments. We report that all observed properties---NSBH merger rate, eccentricity, and spin-orbit misalignment---are explained by the high prevalence of massive stellar triples in the field. Modelling their evolution from the zero-age-main-sequence, we investigate NSBH mergers caused by gravitational perturbations from a tertiary companion. The NS formation decisively impacts the triple stability, preferentially leaving behind surviving NSBHs in compact triple architectures \new{with mild hierarchies}. The rich three-body dynamics of compact, unequal-mass triples enables mergers across a wide range of orbital parameters and provides a natural explanation for an abundance of residual eccentricity and spin-orbit misalignment. We infer a total NSBH merger rate of $\mathcal{R}_{\rm NSBH}\sim1$ -- $23\,\rm Gpc^{-3}\,yr^{-1}$ \new{(within uncertainties on NS kicks)} with more than a few $10\,\%$ exhibiting residual eccentricity $e_{20}>0.1$ or large spin-orbit misalignment $\cos\theta_{\rm BH}<0$, consistent with current observations. The mergers closely track the cosmic star formation rate due to short delay times ($\sim10$ -- $100\,\rm Myr$), include a substantial fraction of burst-like highly eccentric systems ($e_{20} > 0.9$), and almost universally retain eccentricities $e_{20} > 10^{-3}$ detectable by next-generation detectors. If evidence for eccentric and misaligned events solidifies, our results suggest that triple dynamics is the dominant formation channel of gravitational-wave events from NSBH mergers.
\end{abstract}

%% You will be asked to selected these concepts during the sumathbfission process


\section{Introduction} 
Three of the \new{seven} potential neutron star-black hole (NSBH) mergers discovered through gravitational wave (GW) radiation \new{\citep{GW190814,Abbott2021,Abbott2023,GW230529,LIGO-04-new-NSBH}} exhibit enigmatic properties that challenge the standard formation scenario from the evolution of interacting isolated massive binary stars \citep[e.g.,][]{Mapelli2018,Broekgaarden2021,Drozda2022,Xing2024}. On the one hand, \citet{Abbott2021} report an 88\,\% probability that the black hole (BH) spin of GW200115 is misaligned to the binary orbital angular momentum vector by more than $90\,{\rm deg}$. Moreover, the BH spin of the NSBH candidate GW230529 shows strong evidence of being misaligned, but it cannot be ruled out that the event is a binary neutron star (NS) merger \citep{GW230529}. If NSBH mergers result from the common-envelope evolution of isolated binary stars, \citet{Fragione2021} showed that such large spin-orbit tilts could only be accommodated with large NS natal kicks ($\gtrsim150\,\rm km\,s^{-1}$) and high efficiencies for common-envelope ejection ($\alpha_{\rm CE}\gtrsim3$). On the other hand, \citet{Morras2025} found  evidence that the NSBH merger GW200105 \citep{Abbott2021} retained a residual eccentricity $\gtrsim0.1$ upon merging (excluding zero at more than 99\,\% confidence), \new{which aligns with earlier findings in the context of tests for alternative theories of gravity \citep{Fei2024} and which} was later corroborated by \citet{planas2025}, \new{ \citet{Kacanja2025}, and \citet{Jan2025}}. Any measurable residual eccentricity close to merger would imply that the binary must have had an extremely large eccentricity in the past because of highly efficient orbital circularisation by GW radiation \citep{Peters1964}. This would be wholly inconsistent with a formation from isolated massive binary stars \citep[e.g.,][]{Belczynski2002,Fumagalli2024}.

Inferring a non-zero spin-orbit tilt and eccentricity in the aforementioned GW events is sensitive to assumptions of the parameter estimation. Regarding GW200115, \citet{Mandel_2021} use interacting binary star models to advocate for a BH spin prior limited to small magnitudes (as opposed to larger magnitudes allowed by \citet{Abbott2021}) leading to no support for significant spin-orbit misalignment. Meanwhile, \citet{Morras2025} find non-zero eccentricity of GW200105 for all of their chosen priors of the eccentricity except for log-uniform priors with a lower prior bound of $<10^{-4}$. Moreover, GW200105 was effectively a single-detector event recorded by LIGO Livingston while LIGO Hanford was not operational and Virgo yielded a low signal-to-noise ratio \citep{Abbott2021}. However, if evidence for large spin-orbit tilts and non-zero eccentricity becomes more robust in the growing population of detected NSBH mergers, it calls for alternative formation scenarios where the mergers are caused by some dynamical perturbation from the environment.% in order to recover the apparent large fraction of NSBH mergers with spin-orbit misalignment and residual eccentricity. 

In the literature, there exist four main formation channels in which NSBHs (as well as stellar binary BHs and binary NSs) could be dynamically driven to merge and which can generally recover a considerable fraction of non-zero eccentricity and spin-orbit misalignment. First, merging binaries can result from dynamical interactions in dense star clusters. However, multiple works found the rate of NSBH mergers in star clusters to be several orders of magnitude smaller than the estimated rate \new{$\mathcal{R}_{\rm LVK}=9.1$~--~$84\,\rm Gpc^{-3}\,yr^{-1}$ \citep{LIGO-O4-pop}} from GW detections \citep{Clausen2013,Bae2014,Petrovich2017,Belczynski2018,Ye2019,Hoang2020,ArcaSedda2020,Fragione2020,Ye2020}. This is because BHs and NSs are expected to spatially separate in clusters. The heavier BHs mass-segregate into the dense cluster core where their dynamical interactions act as a heating source (``BH burning") that keeps the lighter NSs at larger cluster radii. Second, ultra-wide binaries (with semi-major axis $\gtrsim10^3\,\rm AU$) can be driven to merger by torques exerted by their host galaxy and fly-bys of surrounding stars \citep{Michaely2019,Raveh2022,Stegmann2024}. \citet{Michaely2022} found that the rate of NSBH mergers from wide binaries agrees with the observationally inferred rate if NSs and BHs receive zero natal kicks, which is uncertain. If NSs even receive moderate natal kicks $\gtrsim\mathcal{O}(1)\,\rm km\,s^{-1}$ the vast majority of the loosely bound ultra-wide binaries gets disrupted and the rate of NSBH mergers would be significantly suppressed \citep{Stegmann2024}. Third, NSBHs can merge in active galactic nuclei at high rates, but the details of this model are highly uncertain \citep{Yang2020,McKernan2020,Tagawa2021}. 

In this work, we investigate the formation of NSBH mergers through the fourth channel, where compact object mergers are caused by the presence of a distant bound tertiary companion. This scenario is well-motivated by observations showing that most massive progenitor stars of NSs and BHs are found in hierarchical triples \citep[][]{Moe2017}. In these systems the gravitational perturbation from the tertiary companion can induce long-term, large-amplitude von Zeipel-Kozai-Lidov (ZKL) eccentricity oscillations of the inner binary \citep{Zeipel1910,Kozai1962,Lidov1962} and promote a GW-driven inspiral at close periapsis passage \citep{Ford2000,Wen2003,Antonini2014,Antognini2014,Antonini2016,Antonini2017,Liu2019,Martinez2020,Fragione2020,Bartos2023,Vigna2025}. \citet{FragioneLoeb2019a,FragioneLoeb2019b} focussed on the formation of NSBH mergers from the evolution of the subset of triples with \textit{wide} non-interacting inner binary stars (separated by $\ge10\,\rm AU$). While they did find a significant fraction of tertiary-driven NSBH mergers with finite eccentricity and spin-orbit misalignment, the wide orbits of their triples get easily disrupted by NS natal kicks. Thus, the resulting merger rates are several orders of magnitude lower than the observationally inferred rate unless zero kicks are assumed. Similarly, \citet{Trani2022} considered the evolution of relatively wide compact object triples obtained from direct $N$-body simulations of dissolving young star clusters. They also recovered a considerable fraction of eccentric tertiary-driven NSBH mergers, albeit with a total merger rate that is one to three orders of magnitude lower than the observationally inferred range. In contrast, \citet{Hamers2019,Stegmann2022} did include the high close binary fraction observed for massive stars \citep[e.g.,][]{Sana2012} which leads to a much larger number of surviving systems and NSBH merger rates consistent with the detections. However, they did not fully model the three-body dynamics after NSBH formation nor investigated the detailed properties of any merger, leaving the characteristics of the bulk of NSBH mergers from triples largely unexplored and a comparison to GW data pending.

Here, we investigate the latter and study the formation and properties of NSBH mergers from a realistic population of massive triple stars.

\label{sec:intro}
\begin{figure}
    \centering
    \includegraphics[width=0.8\linewidth]{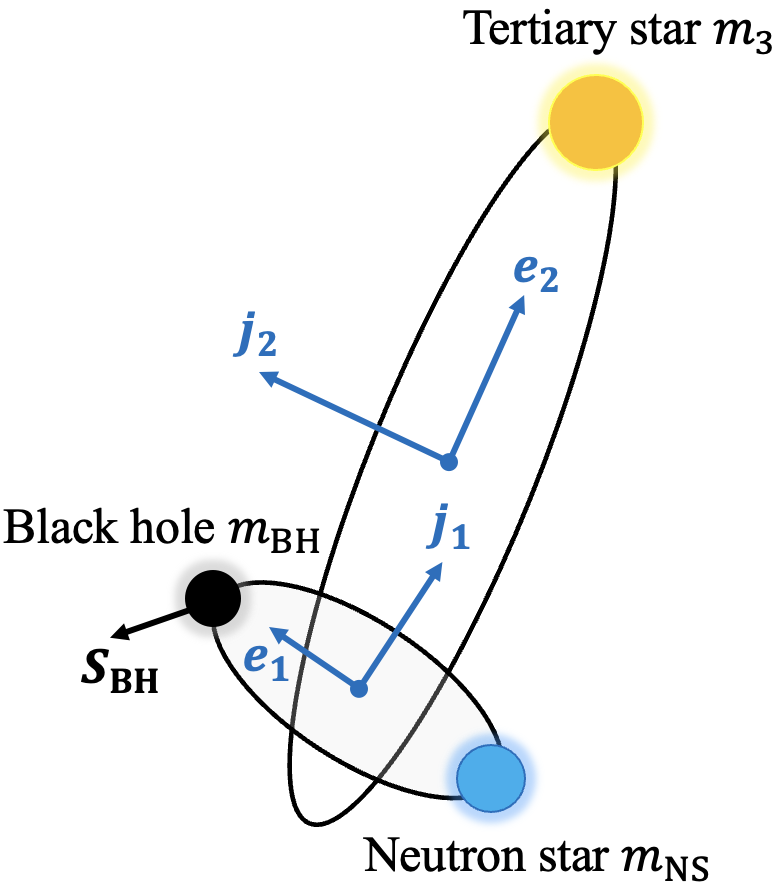}
    \caption{Cartoon of hierarchical NSBH triples whose formation and evolution is studied in this work. The two inner binary stars form an NSBH with masses $m_1=m_{\rm BH}$ and $m_2=m_{\rm NS}$ which is orbited by a distant tertiary companion star with mass $m_3$. The inner and outer orbits are ellipses (denoted by indices ``1" and ``2", respectively) that can be characterised by eccentricity vectors $\mathbf{e_1}$ and $\mathbf{e_2}$ that point towards the orbital periapses and have magnitudes equal to the orbital eccentricities $|\mathbf{e_1}|=e_1$ and $|\mathbf{e_2}|=e_2$, and by dimensionless orbital angular momentum vectors $\mathbf{j_1}$ and $\mathbf{j_2}$ which are perpendicular to the orbital planes \citep[][]{Tremaine2009}. The two semi-major axes of the inner and outer orbits are hierarchical in the sense that the semi-major axes satisfy $a_1\ll a_2$. We also consider the evolution of the BH spin vector $\mathbf{S}_{\rm BH}$ and its angle $\theta_{\rm BH}$ with respect to $\mathbf{j_1}$.}
    \label{fig:sketch}
\end{figure}
\begin{figure*}[ht!]
    \centering
    \includegraphics[width=1\linewidth]{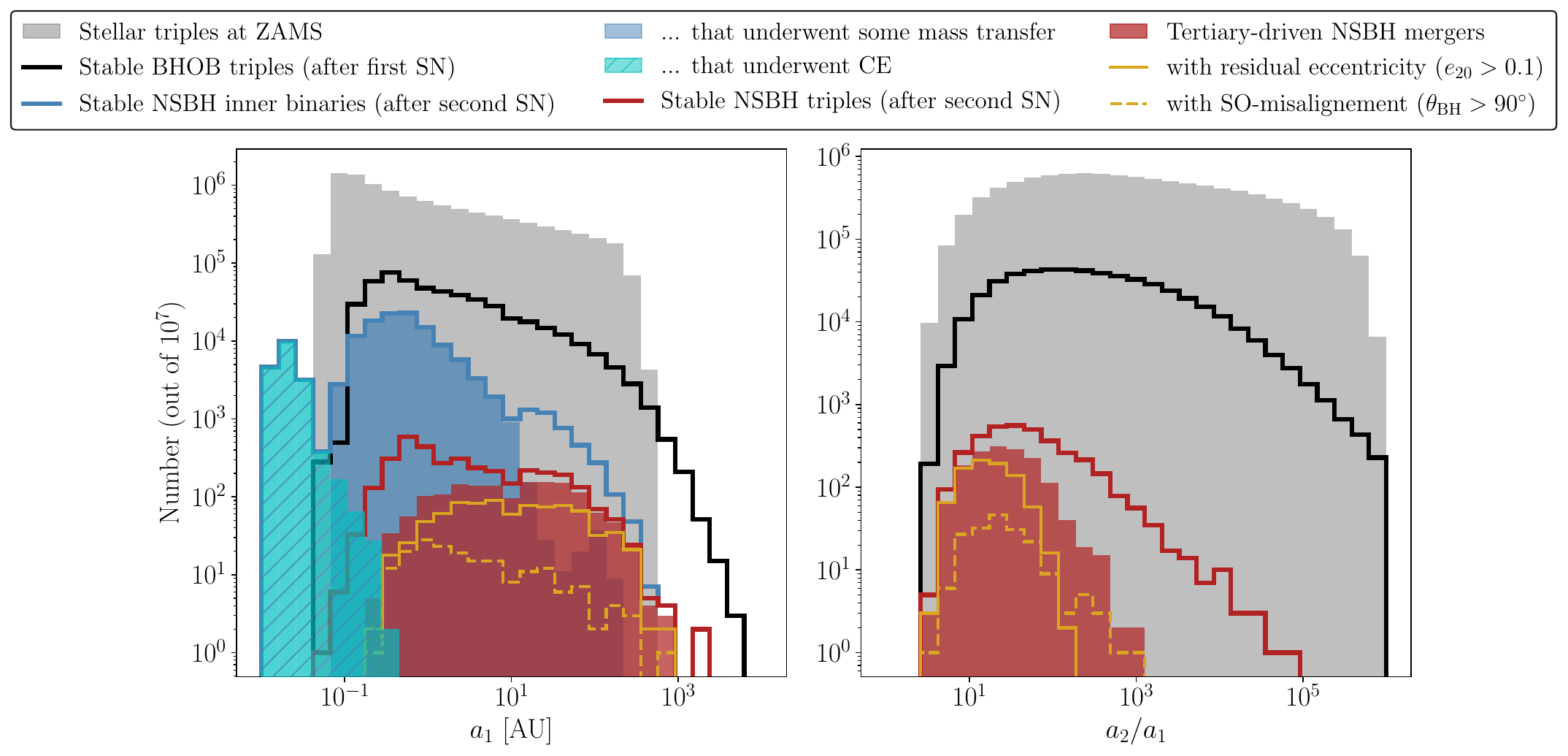}
    \caption{Distributions of the inner semi-major axis (left panel) and semi-major axis ratio (right panel) of triples studied in this work. Grey patches indicate the distributions of $N_{\rm tot}=10^7$ stellar triples at ZAMS (initial conditions), black lines show all triples which survive the first SN and form a BHOB in the inner binary, blue lines show all inner binaries which survive the second SN and form an NSBH (regardless of whether the outer binary remained stable or not), and blue and turquoise patches highlight the subset of NSBHs that underwent some sort of mass transfer and a common-envelope evolution, respectively. Red lines show all stable triples which develop an NSBH in the inner binary, red patches show those that eventually lead to an NSBH merger, and gold solid and dashed lines highlight the subset of those which retain residual eccentricity upon merging ($e_{20}>0.1$) and coalesce with large BH spin-orbit misalignment ($\cos \theta_{\rm BH}<0$), respectively.}
    \label{fig:SMA-histo}
\end{figure*}
\begin{figure*}
    \centering
    \includegraphics[width=.8\linewidth]{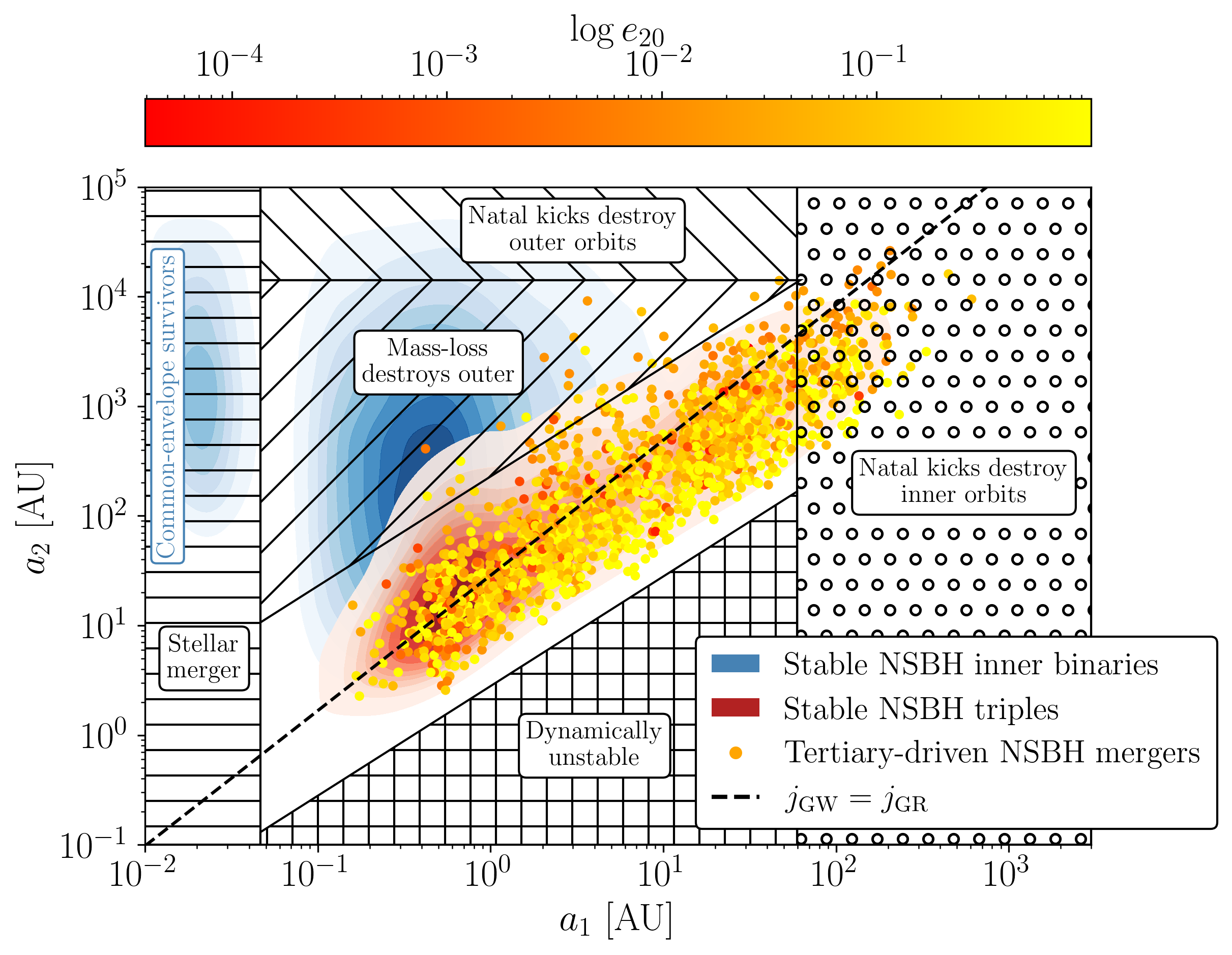}
    \caption{Parameter space of BHOB triples before the second SN. Blue contours show the density of triples with inner binaries that survive the second SN and form stable NSBH binaries. Red contours represent systems successfully forming stable NSBH triples. Scatter points highlight systems which will lead to tertiary-driven NSBH mergers with the colour indicating their residual eccentricity $e_{20}$ at a dominant GW frequency $f_{\rm GW}=20\,\rm Hz$ \citep{Wen2003}. The parameter window for surviving NSBH triples (red) can be understood as a result of several constraining limits, marked with hatched and dotted regions: Diagonally hatched and dotted regions indicate regions where the natal kick tends to disrupt the inner binary, the natal kick tends to disrupt the outer binary, and the SN mass-loss disrupts the outer binary, respectively (see Eqs.~\eqref{eq:a1}~--~\eqref{eq:a1a2}). Below $a_1\approx10\,\rm R_\odot$ inner binaries would undergo a stellar merger at ZAMS (horizontally hatched), and may only be occupied by systems surviving a common-envelope phase. The square hatched region indicates region of dynamical instability of the triple. The black dashed line shows $j_{\rm GW}=j_{\rm GR}$ for a fiducial triple with $m_1=10\,\rm M_\odot$, $m_2=1.5\,\rm M_\odot$, $m_3=1\,\rm M_\odot$, and $e_2=0.67$, respectively.}
    \label{fig:regimes}
\end{figure*}
\section{Methods}

% Sketch: \\
% \begin{itemize}
%     \item Explain what it is that we model. E.g. We consider the evolution of BH + OB systems with tertiary companions in hierarchical triples... As we explain below, we consider triples in which the inner binaries evolve essentially in isolation until the NSBH stage --> throw away all other. 
%     % \item Initial conditions first (e.g. m3 < X, so that it remains a MS star). Resulting triple fraction. 
%     \item Single star evolution (stellar tracks, etc)
%     \item the binary effects that are accounted for (mass transfer, wind, stability, core mass that remains, CE.. the rapid model for interacting binaries. He star mass loss)
%     \item comment on a numerical approach? e.g. grid of masses, mass ratios, periods. possible todo: some plot for the appendix to demonstrate it works. 
%     \item BH vs NS formation. 
% \end{itemize}

We consider the evolution of hierarchical triples composed of an inner binary of two massive stars ($m_1 \geq 10\Msun$, $m_2 \geq 5\Msun$) and a less massive tertiary star ($m_3 < 5\Msun$). Our goal is to model the formation of stable triples that harbour an NSBH in the inner binary (``NSBH triple" hereafter, see Fig.~\ref{fig:sketch}) which is subsequently driven to merge by the gravitational perturbation from the tertiary companion. 
Because the stellar lifetime rapidly decreases with mass ($t \propto M^{-\alpha}$, where $\alpha \approx 2.5-3.5$), the NSBH will form before the tertiary leaves the Main Sequence (MS). Combined with negligible stellar winds of MS stars below $5\Msun$, this allows us to ignore the stellar evolution of the tertiary and treat it as a point mass. During the stellar evolution from the MS to the possible formation of NSBHs, we employ a model for binary star evolution on the inner orbit (described in Sec.~\ref{sec:binary-evolution}), apply the effect of binary mass-loss and any supernova to the orbits of a triple (Sec.~\ref{sec:SN}), and evolve the gravitational three-body dynamics after the NSBH has formed (Sec.~\ref{sec:gravity}). %Not accounting for the gravitational three-body dynamics prior to the NSBH formation neglects the possibility that the inner binary star is driven to some interaction because of ZKL oscillations \citep{Hamers2019,Toonen2020,Stegmann2022,Stegmann2022a}. While in some of the parameter space ZKL oscillations are expected to be quenched by tidal interactions between the inner binary stars, we estimate in Sec.~\ref{sec:tides} the importance of our simplification.
Not accounting for the gravitational three-body dynamics prior to the NSBH formation neglects the possibility that the inner binary star is driven to some interaction because of ZKL oscillations \citep{Hamers2019,Toonen2020,Stegmann2022,Stegmann2022a}, although for most systems we expect them to be quenched by tides between the close inner binary stars.

\subsection{Evolution of inner binaries}
\label{sec:binary-evolution}

For the vast majority of binaries that evolve to become NSBHs \citep[$\gtrsim 95\,\%$;][]{Broekgaarden2021,Xing2024,Liotine2025}, the BH is expected to form first from the more massive primary star ($m_1$), while the progenitor of the NS is still on the MS ($m_2$). This implies an intermediate evolutionary stage of an inner binary with a BH and an OB-type stellar companion (``BHOB triple"). We thus construct a two-stage binary evolution model for the inner binaries, as follows.

\begin{enumerate}[label=\textbf{Stage \Roman*)}, leftmargin=*, align=left]
    \item MSMS $\rightarrow$ BHOB, \label{item:stage-I}
    \item BHOB $\rightarrow$ NSBH. \label{item:stage-II}
\end{enumerate}

Our methodology is similar to rapid grid-based binary evolution codes such as COMBINE \citep{Kruckow2018} or SEVN \citep{Iorio2023}. We use single stellar tracks from \citet{Klencki2020,Klencki2025} to obtain the basic stellar properties (mass, radius, luminosity, core mass, envelope binding energy) as a function of age. The tracks encompass 50 masses, logarithmically spaced from 5 to 100$\Msun$, and we use interpolation to fill in for any given initial mass in that range. We choose a subsolar metallicity $Z = 0.0017 \approx 0.1\Zsun$. Unlike for binary BH mergers, the formation of NSBH systems is found to weakly depend on metallicity \citep{Klencki2018,Chruslinska2019,Broekgaarden2021,vanSon2025}. For both \ref{item:stage-I} and \ref{item:stage-II}, we integrate the evolution of the more massive star until core collapse and BH or NS formation, respectively. The timestep is chosen such that the relative change in mass, radius, and envelope binding energy is not greater than 0.1\,\%. For simplicity, we assume that the inner binary is always formed on a circular orbit. This is justified because the vast majority of NSBH mergers we predict to originate from triples in which the inner binary undergoes mass transfer already before the BH formation, which is expected to quickly circularise the orbit. We account for the widening of the orbits (inner and outer) due to mass loss via winds. We neglect the weak stellar winds from the low mass ($m_3< 5\Msun$) tertiary companion star. At any time during the evolution we check for the dynamical stability of the triples \citep{Mardling} and discard any system which becomes unstable.

If the orbit is sufficiently close (with semi-major axis $a_1\lesssim10\,\rm AU$), the expanding primary may at some point overfills its Roche lobe and initiates mass transfer (MT) onto the inner binary companion. To assess its stability, we first check whether the donor has an outer convective or radiative envelope based on the criteria from \citet{Klencki2020}. For convective donors, we assume that the MT is stable for $q > q_{\rm crit;conv}$, where $q = M_{\rm accretor} / M_{\rm donor}$ at the onset of Roche-lobe overflow (RLOF), and set $q_{\rm crit;conv} = 0.8$ \citep{Hjellming1987,Temmink2023}. For radiative donors, we apply two stability requirements. First, for both \ref{item:stage-I} and \ref{item:stage-II}, we require that for the MT to remain stable, the final post-interaction orbit must be wider than $20\Rsun$ \citep[based on the separation limit found in][]{Klencki2025}. Second, for stellar accretors (i.e., only \ref{item:stage-I}), we require $q > q_{\rm crit;rad}$, where $q_{\rm crit;rad}$ depends on orbital period at RLOF:
\begin{equation}
q_{\rm crit;rad} = \begin{cases} 0.35 & \text{for} \, P_{\rm RLOF} \leq 10 {\rm \, days}\\ q({\rm log}P_{\rm RLOF}/{\rm days}) & \text{for} \, P_{\rm RLOF} > 10 {\rm \, days}
\end{cases},
\end{equation}
where $q({\rm log}P_{\rm RLOF}/{\rm days})$ is a linear fit to $q = 0.35$ at ${\rm log}P_{\rm RLOF}/{\rm days} = 1$ and $q = 0.6$ at ${\rm log}P_{\rm RLOF}/{\rm days} = 3$. This criteria approximately reproduces the MT stability from massive radiative donors found in the COMBINE code for our choice of low accretion efficiency $\beta = 0.1$ \citep[see Figs.~E.6~--~E.9 of][]{Schurmann2025_combine}, based on a detailed study by \citet{Schurmann2024_stability}. The origin of MT instability here is a runaway expansion of the mass gainer star when the accretion rate exceeds a certain threshold \citep[see also][]{Henneco2024,Lau2024}. For a larger accretion efficiency, the mass transfer would be less stable, particularly for periods $P_{\rm RLOF} > 100$ days \citep{Schurmann2025_combine}. We assume that unstable MT with a radiative donor always leads to a stellar merger \citep{Podsiadlowski2001,Kruckow2016,Klencki2021,Marchant2021}. For unstable MT with a convective donor, we consider common envelope (CE) evolution.
For \ref{item:stage-I}, we apply the alpha-lambda energy formalism \citep{vdHeuvel1976,Webbink1984} to determine whether the system survives the CE phase as well as the post-CE orbit. We assume $\alpha_{\rm CE} = 1$ and include 100\,\% contribution from internal and recombination energy into the envelope binding energy calculation \citep[for details see][]{Klencki2021}. For \ref{item:stage-II}, we assume that all CE events lead to surviving binaries (BH + He-star systems) with a post-CE semi-major axis $a_{\rm post-CE} = 4 \Rsun$, i.e., sufficiently close to produce a GW-driven NSBH merger even without the aid of the tertiary companion. 
% We keep $a_{\rm post-CE}$ as a free parameter and consider three values: $4.0$, $8.0$, or $12.0\Rsun$.
Our choice to assume 100\,\% survival rate of CE events for \ref{item:stage-II} allows us to gauge the upper limit for the contribution of the CE channel to NSBH mergers. The choice to fix $a_{\rm post-CE}$ is motivated by the fact that the exact final separation of post-CE systems is poorly known as it is likely determined by the complex interaction between the binary orbit and the circumbinary matter ejected during the dynamical phase of CE \citep[e.g.][]{Gagnier2023,Wei2024}. Moreover, in this work they merely serve as a reference to compare our tertiary-driven NSBH mergers to a model where NSBHs could even merge in isolation. We note that a difference choice of $a_{\rm post-CE}$ would result in a somewhat different distribution of BH spin-orbit misalignments from the CE channel: the wider the orbit, the easier it is to tilt by the natal kick associated with NS formation, see Sec.~\ref{sec:SN} \citep{Fragione2021}.

For mass transfer events that remain stable, we follow the method detailed in Sec.~2 and App.~A of \citet{Klencki2025}. Briefly, we calculate the change in orbital semi-major axis as:
\begin{equation}
    a_{\rm fin} = a_{\rm ini} + \int_{M_{\rm don;ini}}^{M_{\rm don;fin}} \frac{{\rm d}a}{{\rm d}M_{\rm don}} \left( M_{\rm acc}, \beta, \gamma \right)    {\rm d}M_{\rm don} \, \, ,
\label{eq.SMT_integral}
\end{equation}
where $a_{\rm ini}$ ($a_{\rm fin}$) is the semi-major axis before (after) mass transfer and $M_{\rm don}$ ($M_{\rm acc}$) is the mass of the donor (accretor). For the accretion efficiency we set $\beta = 0.1$ for \ref{item:stage-I} and $\beta = 0.0$ for \ref{item:stage-II}. The first choice roughly approximates results from detailed models with rotationally-limited accretion (e.g. \citealt{Langer2003,Renzo2021}, although $\beta > 0.1$ may be required for some systems, see  \citealt{Vinciguerra2020,Lechien2025}), the second is motivated by the Eddington accretion limit. We assume that the non-accreted mass is ejected from the proximity of the accretor ($\gamma = M_{\rm don}/M_{\rm acc}$). 
For case B/C mass transfer, we assume that the donor transfers 90\,\% of its envelope ($M_{\rm don;fin} = M_{\rm He;core} + 0.1 M_{\rm env}$). For case A, this amount is less certain due to the receding convective core \citep{Shikauchi2024} and we calculate $M_{\rm don;fin}$ based on analytical fits to detailed binary evolution models provided by \citet[][their SMC variant]{Schurmann2024_caseA}. Once a helium star forms as a result of mass transfer, we estimate its further mass loss by applying the wind from \citet{Hamann1995} lowered by a factor of 10 \citep{Yoon2006}.

In order to assess the final fate of a star, we apply the birth function for BH and NS formation in close binaries from \citet{Woosley2020}. As their results are based on naked helium star models, we map their pre-supernova masses to our final helium core masses. In addition, following \citet{Hurley2002}, we assume that stars with helium cores between $1.6$ and $2.3\Msun$ form a NS through an electron-capture supernova (ECSN). We note that \citet{Podsiadlowski2004} argued for a wider mass range for ECSN for stars evolving in close binaries, whereas \citet{Willcox2021} showed that a narrowed mass range for ECSN from single stars might be required to match the number of low-velocity pulsars.  Regardless of the SN type, we assume that all NSs form with a mass of $m_{\rm NS}=1.5\Msun$. For BH formation, if the collapsing star still retains some of its hydrogen envelope, we assume that 50\,\% of this envelope is accreted onto the BH and the rest is ejected. Guided by \citet{Woosley2020}, we assume that naked helium stars collapse directly to form BHs with only $5\,\%$ of their mass lost in neutrinos.
Besides the mass loss, we assume weak BH natal kicks drawn from a Maxwellian distribution with velocity dispersion $\sigma_k = 5\,\rm  {km\,s^{-1}}$. This is consistent with theoretical predictions for neutrino-driven kicks \citep{Janka2024} which struggle to produce kicks $> 100 \,\rm{km\,s^{-1}}$ even for large neutrino emission asymmetries \citep{Burrows2024}, as is supported by the recently discovered BH binary VFTS 243 \citep{VignaGomez2024}. The NS natal kicks are expected to be larger and play a significant role in the survival of triples. We therefore consider several model variations for the NS kick magnitudes $v_k$: 

\begin{enumerate}[label=\alph*)]
    \item Maxwellian\_265\_30: Maxwellian with $\sigma_k = 265 \rm \, {km\,s^{-1}}$ \citep{Hobbs2005}. For ECSN, we instead apply $\sigma_k = 30 \rm \, {km\,s^{-1}}$ \citep{Jones2016,Gessner2018}.  
    \item Maxwellian\_100\_10: Maxwellian with $\sigma_k = 100 \rm \, {km\,s^{-1}}$. For ECSN, we instead apply $\sigma_k = 10 \rm \, {km\,s^{-1}}$.
    \item Igoshev: Maxwellian with $\sigma_k = 336 \rm \, {km\,s^{-1}}$ in 80\% of cases and $\sigma_k = 45 \rm \, {km\,s^{-1}}$ 20\% otherwise, randomly drawn \citep{Igoshev2021}, see also \citep{Verbunt2017}. No separate ECSN kick treatment. 
    \item Valli: Maxwellian with $\sigma_k = 100 \rm \, {km\,s^{-1}}$ for stars that underwent mass transfer, otherwise $\sigma_k = 265 \rm \, {km\,s^{-1}}$. This is motivated by \citet[][]{Valli2025} but is only a rough approximation: in particular we do not include their model for anisotropic kicks (see Sec.~\ref{sec:SN}). For ECSN, we instead apply a Maxwellian with $\sigma_k = 10 \rm \, {km\,s^{-1}}$, regardless of mass transfer history.\footnote{The assumption of $\sigma_k = 10 \rm \, {km\,s^{-1}}$ to all ECSN may not be a good description of the results from \citet[][]{Valli2025}.}
    \item Disberg: log-normal distribution with $\mu = 5.6$ and $\sigma = 0.68$ \citep{Disberg2025}. No separate ECSN kick treatment. 
    \item Bray: deterministic kick expressed as $v_k = \alpha (M_{\rm ejecta}/M_{\rm NS}) + \beta$, where $\alpha = 115 \rm \, {km\,s^{-1}}$ $\beta = 10 \rm \, {km\,s^{-1}}$, and $M_{\rm ejecta}$ is the mass lost upon the  NS formation \citep{Bray2018,Richards2023}. No separate ECSN kick treatment. 
    \item Zero: no NS natal kicks. 
\end{enumerate}

\subsection{Impact of supernovae on triple orbits}\label{sec:SN}
Whenever a supernova (SN) occurs in the inner binary the sudden mass loss and natal kicks can change the orbital parameters and possibly disrupt the triple. Briefly, we first compute the pre-SN relative separation and velocity vectors $\mathbf{r}_1$ and $\mathbf{v}_1$ of the inner binary, assuming that the SN occurs at a random mean anomaly. We then apply a random natal kick to the SN remnant by changing the relative velocity vector as
\begin{equation}
    \mathbf{v}_1\rightarrow\mathbf{v}_{1,n}=\mathbf{v}_1+\mathbf{v}_k,
\end{equation} 
where the kick velocity vector $\mathbf{v}_k$ points in a isotropically random direction and has a magnitude drawn from either one of the distribution described in Sec.~\ref{sec:binary-evolution}. Furthermore, the star undergoing SN is subject to sudden mass-loss given by 
\begin{equation}
     m_{1(2)}\rightarrow m_{1(2),n}=m_{1(2)}-\Delta m_{1(2)},
\end{equation}
where $m_{1(2)}>\Delta m_{1(2)}\ge0$. The new mass $m_{1(2),n}$ is either given by the BH mass $m_{\rm BH}$ or the NS mass $m_{\rm NS}$, depending on which star is about to explode. Using the new relative velocity vector and mass we compute the updated orbital parameters and discard any disrupted binary which has a new semi-major axis $a_{1,n}<0$ or eccentricity $e_{1,n}\notin [0,1)$. Generally, large changes of the orbital parameters or disruptions take place if $v_k\gtrsim\mathcal{O}(1)v_1$ which for a given kick magnitude translates into an upper limit on $a_1=a_1(v_1)$ for surviving inner binaries.

The mass-loss and natal kick associated with the SN also imparts a kick to the centre-of-mass velocity of the inner binary which modifies the orbit of the outer binary and potentially disrupts it. We update the outer orbit in an analogue way to the inner binary, where the new relative velocity vector of the outer binary and total mass of the inner binary are given by \citep{Lu2019}\footnote{Different to Eq.~(24) of \citet{Lu2019} we find a negative sign in front of the $\mathbf{v}_k$-term in Eq.~\eqref{eq:new-outer}, which is irrelevant for isotropically random kick directions.}
\begin{widetext}
\begin{equation}
    \mathbf{v}_2\rightarrow\mathbf{v}_{2,n}=\mathbf{v}_2+\frac{m_{1(2)}\Delta m_{2(1)}\mathbf{v}_1}{[m_{1(2)}+m_{2(1),n}][m_{1}+m_{2}]}-\frac{m_{2(1),n}\mathbf{v}_k}{m_{1(2)}+m_{2(1),n}},\label{eq:new-outer}
\end{equation}
\begin{equation}
    m_1+m_2\rightarrow m_{1(2),n}+m_{2(1)},
\end{equation}
\end{widetext}
and discard any disrupted triple which have a new outer binary semi-major axis $a_{2,n}<0$ or eccentricity $e_{2,n}\notin [0,1)$. Eq.~\eqref{eq:new-outer} shows that the SN affects the outer binary in a more complicated way than the inner, which can be understood as follows. The term proportional to the change of remnant momentum $m_{2(1),n}\mathbf{v}_k$ represents the kick imparted to the inner binary centre-of-mass due to the natal kick of the SN remnant. The other term proportional to $\Delta m_{2(1)}\mathbf{v}_1$ describes the recoil imparted to the inner binary centre-of-mass by the ejected mass. Large changes of the outer orbital parameters and disruptions take place if $|\mathbf{v}_{2,n}-\mathbf{v}_2|\gtrsim\mathcal{O}(1)v_2$. Thus, either of the two terms sets a different constrain for the survivability of the triples. On the one hand, requiring $v_2\lesssim{m_{2(1),n}v_k}/[{m_{1(2)}+m_{2(1),n}}]$ results in an upper limit on $a_2$, i.e., the outer orbit must be close enough to survive a natal kick of a given magnitude. On the other hand, requiring $\mathbf{v}_2$ to be less than the $\mathbf{v}_1$-term in Eq.~\eqref{eq:new-outer} results in a upper limit on $a_2/a_1$. This is because magnitude of $\mathbf{v}_1$ grows for smaller $a_1$ by Kepler's law. Thus, the velocity of the mass ejecta and the recoil imparted to the inner binary centre-of-mass will be generally larger for smaller $a_1$, which in turn reduces the maximum $a_2$ that a surviving triple could have. In summary, requiring $v_k\lesssim v_1$ and the $v_{1(r)}$-terms to be smaller than $v_2$ we obtain three constrains beyond which the inner and outer binary, respectively, undergo large orbital changes and potentially get disrupted,
\begin{align}
    a_1&\lesssim\frac{G(m_1+m_2)}{v_k^2},\label{eq:a1}\\
    a_2&\lesssim\frac{G(m_1+m_2+m_3)(m_{1(2)}+m_{2(1),n})^2}{m_{2(1),n}^2v_k^2},\label{eq:a2}\\
    \frac{a_2}{a_1}&\lesssim\frac{(m_1+m_2+m_3)(m_1+m_2)[m_{1(2)}+m_{2(1),n}]^2}{m_{1(2)}^2\Delta m_{2(1)}^2}.\label{eq:a1a2}
\end{align}
In Sec.~\ref{sec:results}, we show that for the NS formation these constrains critically shape the distributions of surviving triples with an NSBH in the inner binary and their long-term dynamical evolution towards becoming a GW source.

\subsection{Gravitational three-body dynamics}
\label{sec:gravity}
For any triple that successfully formed an NSBH in the inner binary without being disrupted or becoming dynamically unstable, we integrate the long-term secular three-body dynamics as follows. From the post-SN orbital parameters we construct the eccentricity vectors $\mathbf{e}_{1(2)}$ and dimensionless orbital angular momentum vectors $\mathbf{j}_{1(2)}$ of the inner and outer orbit, respectively (see Fig.~\ref{fig:sketch}). The eccentricity vectors are defined to point towards the orbital periapsis and have lengths $|\mathbf{e}_{1(2)}|={e}_{1(2)}$. The dimensionless orbital angular momentum vectors $\mathbf{j}_{1(2)}$ are perpendicular to the orbital plane of the inner and outer binary, respectively, and have lengths $|\mathbf{j}_{1(2)}|=\sqrt{1-{e}_{1(2)}^2}$. Additionally, we consider the spin vector of the BH $\mathbf{S}_{\rm BH}=\mathbf{\chi}_{\rm BH}Gm_{\rm BH}^2/c$, where we assume a dimensionless spin parameter $\mathbf{\chi}_{\rm BH}=0.25$ consistent with the low spin magnitudes inferred from previous BH merger events \citep{Abbott2023} and we assume that $\mathbf{S}_{\rm BH}$ is initially aligned with the pre-SN inner orbital angular momentum, as a result of stellar evolution. We consider the equations of motion for $\mathbf{e}_{1(2)}$, $\mathbf{j}_{1(2)}$, $\mathbf{S}_{\rm BH}$, and $a_1$ given by the ZKL effect to the octupole order accounting for the Newtonian dynamics of hierarchical three-body systems, GW radiation and Schwarzschild precession of the inner binary, and relativistic spin-orbit coupling. That is, we include Eqs.~(17)~--~(20) of \citet{Liu2015} and Eqs.~(13), (15)~--~(17), and~(25)~--~(26) of \citet{Rodriguez2018}, which are lengthy and not repeated here for convenience. We integrate them over time using the Radau method \citep{Radau} with relative and absolute tolerances of $10^{-10}$. The integrations are terminated if either the NSBHs merge or if the integration time exceeds $\min(10^3\times t_{\rm ZKL},13\,\rm Gyr)$, where the ZKL timescale of the leading-order quadrupole effect is given by
\begin{equation}
    t_{\text{ZKL}} = \frac{m_{\rm BH} + m_{\rm NS}}{m_3 \nu} \left( \frac{a_2}{a_1} \right)^3 \left(1 - e_1^2 \right)^{3/2},\label{eq:tLK}
\end{equation}
with $\nu = \sqrt{{G (m_{\rm BH} + m_{\rm NS})}/{a_1^3}}$.

Near a large eccentricity excitation of the inner binary the timescale for changes in the orbital elements could become shorter than the outer or inner orbital period, rendering the secular equations of motion inaccurate and likely underestimating the number of eccentric mergers by a few ten percent compared to direct three-body integration \citep[e.g.,][]{Antonini2014,Grishin2018,Hamilton2024b}. Moreover, we do not attempt to model tidal deformation and mass-transfer of NSs which may alter the properties of NSBHs near the merger \citep{Davies2005,Foucart2024,Zenati2025}, which is beyond the scope of this work. \new{Truncating the secular equations of motion at the octupole order we also neglect any higher-order terms which may become relevant for driving large eccentricities \citep{Will2017}. Following \citet{Will2021} we consider the ratio between the timescales of octupole effects and the next-order hexadecapole terms}
\begin{equation}
    \new{\frac{T_{\rm oct}}{T_{\rm hex}}\sim\frac{a_1}{a_2e_2(1-e_2^2)}\frac{1-3\eta}{\sqrt{1-4\eta}},}
\end{equation}
\new{where $\eta=m_{\rm NS}m_{\rm BH}/(m_{\rm NS}+m_{\rm BH})^2$. Since $m_{\rm NS}\ll m_{\rm BH}$ we find $(1-3\eta)/\sqrt{1-4\eta}\approx1$. Thus, hexadecapole effects may become relevant for mildly hierarchical triples $a_1/a_2\lesssim0.1$ and small outer eccentricities $e_2\lesssim0.1$, which are suppressed by assuming a thermal distribution (see Sec.~\ref{sec:IC}).}

\new{We also evolve the sample of NSBH triples in one model (Maxwellian\_100\_10) with the direct $N$-body integrator {\tt MSTAR} \citep{Rantala2020,Mannerkoski2023}, in order to estimate the impact of secular break-down at large eccentricities \citep{Antonini2014} and truncation at octupole order on our results. {\tt MSTAR} is based on an algorithmically regularized integration technique \citep[e.g.,][]{Mikkola1999,Preto1999,Mikkola2008,Hellstrom2010,Trani2023} and uses the Gragg-Bulirsch-Stoer (GBS) extrapolation technique \citep{Gragg1965,Bulirsch1966} to evolve the equations of motion to 3.5 post-Newtonian order at high precision.}

\new{Using the output of any direct $N$-body integration to extract a physically meaningful eccentricity from the relative separation and velocity vectors $\mathbf{r}_1$ and $\mathbf{v}_1$, respectively, of merging compact object binaries is a non-trivial task. Obtaining the geometric eccentricity $e_g=(r_{\max}-r_{\min})/(r_{\max}+r_{\min})$ as utilised in eccentric waveform analyses \citep[e.g.,][]{Shaikh2023} requires precise numerical methods to accurately resolve the apo- and pericentre $r_{\max}$ and $r_{\min}$, respectively, of the inpiralling binary. Meanwhile, calculating the Keplerian eccentricity form the Laplace–Runge–Lenz vector of the instantaneous vectors $\mathbf{r}_1$ and $\mathbf{v}_1$ (or post-Newtonian variants \citep{Memmesheimer2004,Mannerkoski2019}), as widely done in previous population studies \citep[e.g.,][]{Rodriguez2018b,Zevin2021}, can yield unphysically large eccentricity values if the binary begins to rapidly precess close to merger \citep{Loutrel2019}. Here, we do use the Keplerian eccentricity $e_K$ and semi-major axis $a_K$ calculated from}
\new{
\begin{align}
    \mathbf{e}_K&={\mathbf {v}_1 \times (\mathbf{r}_1\times\mathbf{v}_1)  \over {G(m_{\rm BH}+m_{\rm NS}) }}-{\mathbf {r}_1  \over { {r}_1 }}\\
    \frac{1}{a_K}&=\frac{2}{r_1}-\frac{v_1^2}{G(m_{\rm BH}+m_{\rm NS})}.
\end{align}}
\new{However, for this calculation we use the last $N$-body output ($\mathbf{r}_1,\mathbf{v}_1$) of merging binaries where 
\begin{align}
    r_1&>200\frac{G(m_{\rm BH}+m_{\rm NS})}{c^2} {\quad\rm and}\label{eq:200}\\
    e_K&>\mathcal{P}=\frac{G(m_{\rm BH}+m_{\rm NS})}{c^2}\frac{3-\eta}{a_K(1-e_K^2)}\label{eq:Will}.
\end{align}
Crucially, requiring $e_K>\mathcal{P}$ in Eq.~\eqref{eq:Will} corresponds to Will's condition  ensuring that the binary is still far enough from merging so that $e_K$ correctly traces circularisation upon gravitational-wave emission \citep{Will2019}, while Eq.~\eqref{eq:200} is imposed for numerical reasons as the direct $N$-body output of {\tt MSTAR} can become sparse for separations closer to merger. We then proceed as \citet{Zevin2021} and use $e_K$ and $a_K$ as input to Peter's equations \citep{Peters1964} to obtain the eccentricity $e_{20}$ when the dominant gravitational-wave harmonics has a frequency of 20\,Hz \citep{Wen2003}.}

\subsection{Initial conditions}\label{sec:IC}
On the zero-age-main-sequence (ZAMS) we sample the initial mass of the primary star in the inner binary from an initial mass function $p(m_1)\propto m_1^{-2.3}$ between $10$ and $100\,\rm M_\odot$ \citep{Kroupa}. We follow the observationally inferred distribution of massive binary stars by \citet{Sana2012} and sample the inner binary mass ratio $q_1=m_2/m_1$ from a power-law $p(q_1)\propto q_1^{-0.1}$ and its orbital period from $p(\log P_1/{\rm days})\propto(\log P_1/{\rm days})^{-0.55}$. We reject any inner binary which would overflow its Roche-lobe at ZAMS. We assume that the inner binaries are initially circular ($e_1=0$) and sample the outer binary from a thermal eccentricity distribution, a log-uniform semi-major axis distribution, and the outer binary mass ratio $q_2=m_3/m_1$ from a broken power-law $p(q_2)\propto q_2^{-1.2}$ below $q_2<0.3$ and $\propto q_2^{-2.0}$ above $q_2<0.3$ \citep{Moe2017}, limited to $0.08\,{\rm M_\odot}<m_3<5\,\rm {\rm M_\odot}$. For any given inner binary, we repeat sampling the outer binary until the triple is dynamically stable \citep{Mardling}. We then  discard any triple with an outer semi-major axis above $a_2>5\times10^4\,\rm AU$ because they are likely disrupted by stellar fly-bys and tides from the host galaxy \citep{Jiang,El-Badry2018,Hamilton2024,Stegmann2024}. Finally, the orbital orientations of the inner and outer orbital frame are sampled from random isotropic distributions, leading to a random mutual inclination between the inner and outer orbit. Using the procedure described above, we construct a total number of $N_{\rm tot}=10^7$ stellar triples which we evolve in this work.

\begin{figure*}
    \centering
    \includegraphics[width=1\linewidth]{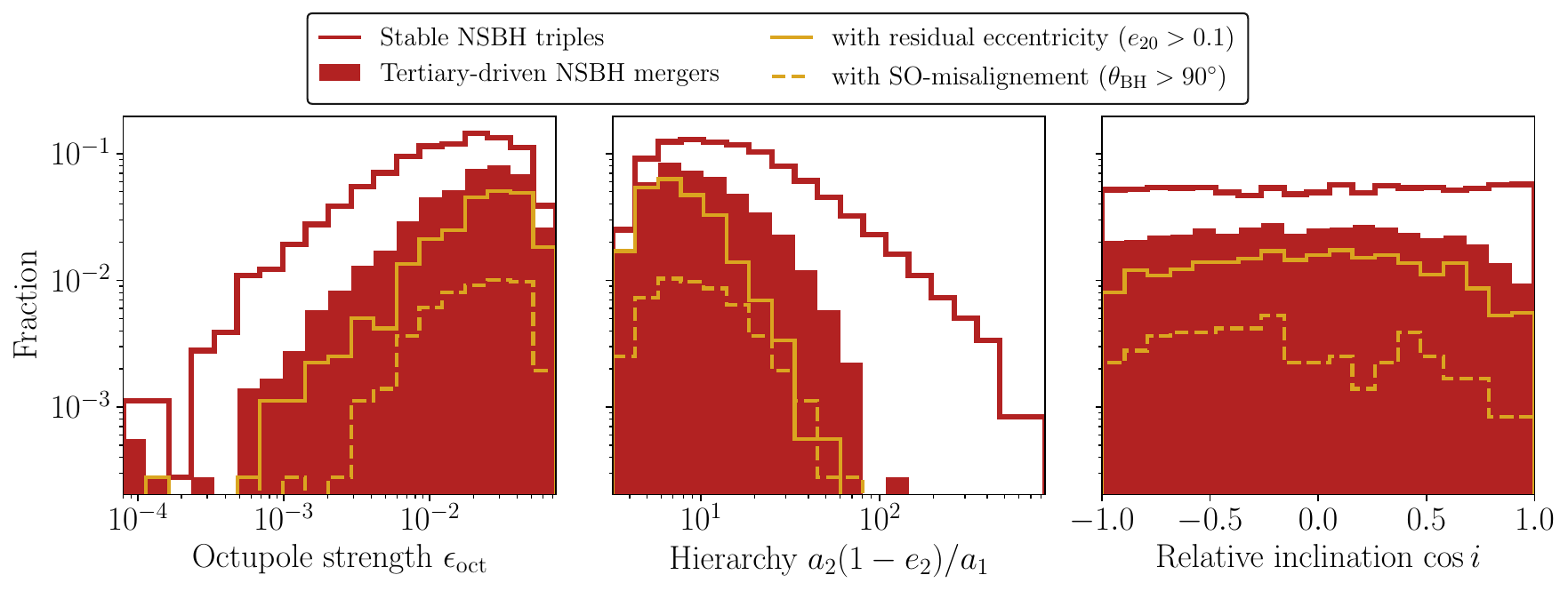}
    \caption{Properties of surviving triples harbouring an NSBH in the inner binary (red line) and of the subset leading to mergers (red patches). NSBHs which retain significant residual eccentricity ($e_{20}>0.1$) upon merging and which coalesce with large BH spin-orbit misalignment ($\theta_{\rm BH}>90\,\rm deg$) are highlighted by solid and dashed golden lines, respectively. All parameters are shown at the time when the NSBHs form (i.e., after the second SN). The fractions on the y-axes are defined with respect to the total number of all massive triples ($10\,{\rm M_\odot}<m_1<100\,{\rm M_\odot}$) at ZAMS.}
    \label{fig:oct}
\end{figure*}
\begin{figure*}
    \centering
    \includegraphics[width=.66\linewidth]{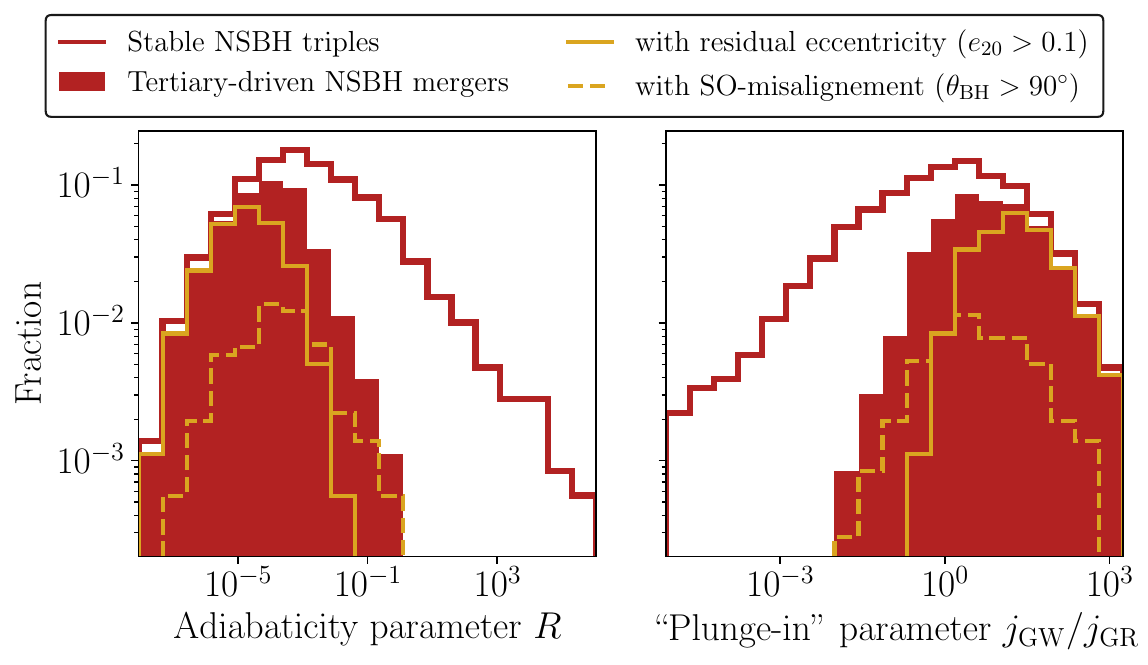}
    \caption{Distribution of the adiabaticity parameter $R=t_{\rm ZKL}/t_{\mathbf{S}_{\rm BH}}$ (see Eqs.~\eqref{eq:tLK} and~\eqref{eq:tS}) and $j_{\rm GW}/j_{\rm GR}$ (Eqs.~\eqref{eq:j_GW} and~\eqref{eq:j_GR}) of NSBH triples. Colours are as in Fig.~\ref{fig:oct}.}
    \label{fig:R_j}
\end{figure*}
\section{Results}\label{sec:results}
Starting on the ZAMS, the orbits of stellar triples have to survive two consecutive SNe and any inner binary mass transfer episode in order to successfully develop stable NSBH triples that could eventually lead to tertiary-driven mergers. In Fig.~\ref{fig:SMA-histo} we show how the properties of surviving systems change across different times during the stellar evolution. %About $\sim5\,\%$ of all systems initiated at ZAMS (grey patches) survive any stable mass transfer episode on the ZAMS and successfully form stable BHOB triples after the first SN (black line). The orbits of their inner binaries slightly expanded as a result of mass transfer and mass-loss associated with the BH formation. About $\sim1\,\%$ successfully form a stable NSBH in the inner binary after surviving 
We find that a few $\sim0.1\,\%$ of all stellar triples initiated at ZAMS (grey patches) survive the stellar evolution and form a stable NSBH triple (red lines) and about $\sim40\,\%$ thereof subsequently lead to a tertiary-driven merger (red patches). The latter fraction is similar to the merger fraction found by \citet{FragioneLoeb2019a} for their population of wider NSBH triples. During stellar evolution the NS formation at second SN is the most detrimental event to the triple stability and critically truncates the parameter distribution of the survivors. On the one hand, the natal kick of the newly formed NS poses the risk to disrupt the inner binary producing a dearth of stable NSBH inner binaries (blue line) above $a_1\gtrsim10^2\,\rm AU$ (left panel). On the other hand, the mass-loss associated with the SN imparts a kick to the inner binary centre-of-mass which causes a steep decline in the distribution of $a_2/a_1$ of surviving NSBH triples above $\gtrsim10^2$ (right panel). The stellar evolutionary constraints are further illustrated in Fig.~\ref{fig:regimes} showing that the surviving NSBH triples occupy a somewhat narrow strip in $a_2$ versus $a_1$ which is predominantly determined by the mass-loss associated with the second SN and the need for the outer orbit to survive it. Consequently, we find different natal kick models only to affect the upper limit on the inner binary semi-major axis of surviving NSBH triples but not substantially changing the semi-major axis ratio distribution, as shown in Fig.~\ref{fig:a1_a1_models}.

%The surviving NSBH triples occupy a distinct region of parameter space that facilitates the formation of tertiary-driven mergers with spin-orbit misalignment and residual eccentricity. On the one hand, the SN kick imparted to the newly formed NS disrupts triples in wide architectures and is only withstood by triples with relatively small inner and outer semi-major axes. This is shown in the two lower left panels of Fig.~\ref{fig:triple-properties} where the inner NSBH semi-major axis distribution of surviving triples peaks at around $a_1\sim1\,\rm AU$ while the ratio $a_2(1-e_2)/a_1$ at $\sim\mathcal{O}(10)$. One the other hand, NSBH triples are characterised by a small inner binary mass ratio $q=m_{\rm NS}/m_{\rm BH}\ll1$ since $m_{\rm NS}\ll m_{\rm BH}$. Therefore, we obtain tertiary-driven NSBH mergers that are, by and large, characterised by \ref{item:oct} strong octupole effects, \ref{item:non-adiabatic} a non-adiabatic BH spin evolution, \ref{item:ecc} a peculiar dynamics in which the NSBH mergers ``plunge" \citep{Rodriguez2018} directly into a highly eccentric, GW-dominated phase without being significantly quenched by Schwarzschild precession, and \ref{item:merger-times} short delay times from ZAMS to merger. Moreover, \ref{item:binary-interactions} the evolution of the NSBH progenitor stars is dominated by close binary interactions, as explained below.
As a result, the surviving NSBH triples naturally populate a distinct region of parameter space that facilitates the formation of tertiary-driven mergers with spin-orbit misalignment and residual eccentricity. On the one hand, the SN kick imparted to the newly formed NS disrupts triples in wide architectures and is only withstood by triples with relatively small inner and outer semi-major axes. One the other hand, NSBH triples are characterised by a small inner binary mass ratio $q=m_{\rm NS}/m_{\rm BH}\ll1$ since $m_{\rm NS}\ll m_{\rm BH}$. Therefore, we obtain tertiary-driven NSBH mergers that are, by and large, characterised by \ref{item:oct} strong octupole effects, \ref{item:non-adiabatic} a non-adiabatic BH spin evolution, \ref{item:ecc} a peculiar dynamics in which the NSBH mergers ``plunge" \citep{Rodriguez2018} directly into a highly eccentric, GW-dominated phase without being significantly quenched by apsidal Schwarzschild precession of the inner binary, and \ref{item:merger-times} short delay times from ZAMS to merger. Moreover, \ref{item:binary-interactions} the evolution of the NSBH progenitor stars is dominated by close stellar binary interactions, as explained below.

%Thus, wide triples with large semi-major axis ratios $a_2/a_1\ggg1$ get disrupted during NS formation and only sufficiently compact three-body configurations survive has significant impact on the subsequent evolution of the NSBH triples. Additionally, NSBH triples are characterised by a small inner binary mass ratio $q=m_{\rm NS}/m_{\rm BH}\ll1$ since $m_{\rm NS}\ll m_{\rm BH}$. Therefore, the surviving triples are in a peculiar dynamical regime in which the NSBH mergers \ref{item:ecc} ``plunge" directly into a highly-eccentric GW-dominated regime without being affected by the Schwarzschild precession, \ref{item:non-adiabatic} evolve non-adiabatically, and \ref{item:oct} are impacted by strong octupole effects, as explained in the following.

\begin{enumerate}[label=(\roman*), leftmargin=*]
    \item\label{item:oct}\textit{Octupole effects enable mergers across a wide range of initial relative inclinations:} The secular equations of motion of hierarchical triples are obtained by expanding the three-body Hamiltonian in semi-major axis ratio $a_1/a_2$ \citep[e.g.,][]{Harrington1968}. To the lowest (``quadrupole") order large inner binary eccentricities are only excited if the relative inclination between the inner and outer orbital planes is near-perpendicular. Large eccentricities across a much wider range of relative inclinations can be excited if the next-order (``octupole") terms becomes relevant \citep[e.g.,][]{Ford2000,Liu2015,Naoz2016} whose relative strength can be characterised as \citep{Naoz2013}
    \begin{equation}
        \epsilon_{\rm oct}=\frac{m_{\rm BH}-m_{\rm NS}}{m_{\rm BH}+m_{\rm NS}}\frac{a_1}{a_2}\frac{e_2}{1-e_2^2}.\label{eq:oct}
    \end{equation}
    As shown in Fig.~\ref{fig:oct}, $\epsilon_{\rm oct}$ of the surviving triples can become as large as $\epsilon_{\rm oct}\lesssim0.1$ (left panel) because the inner binary masses of NSBHs are very unequal, the semi-major axis ratio of surviving NSBH triples is relatively compact, and the thermal eccentricity distribution of the outer binary disfavours circular outer orbits (middle panel). NSBH mergers predominantly occur with nearby tertiaries where $a_2(1-e_2)/a_1\lesssim\mathcal{O}(10)$ and relatively large octupole effects $0.01\lesssim\epsilon_{\rm oct}\lesssim0.1$ \citep{Liu2015}. As a result of the strong octupole effects, mergers are possible across the full range of initial relative inclinations between the inner and orbital plane (right panel) and do not require the somewhat fine-tuned near-perpendicular configuration of the quadrupole-level approximation \citep[e.g., needed for tertiary-driven binary black hole mergers which tend to have larger mass ratios,][]{Antonini2017}.
    %\begin{figure*}
    %\centering
    %\includegraphics[width=1\linewidth]{merger_1576_polished.png}
    %\caption{Evolution of an example tripe that leads to an NSBH merger with large BH spin-orbit misalignment and significant residual eccentricity. The left panel shows large-amplitude ZKL oscillations of the NSBH eccentricity $e_1$ (black line) as a function of time since NSBH formation. Meanwhile, the direction of the NSBH orbital angular momentum oscillates while the BH spin direction remains effectively constant, which leads to large changes of the spin-orbit angle (red line). The system merges after a few cycles at around $t\approx1.4\,\rm Myr$ with a spin-orbit misalignment close to $90\,\rm deg$. The right panel shows the eccentricity evolution as a function of the dominant GW frequency $f_{\rm GW}$.}
    %\label{fig:example}
%\end{figure*}
    \begin{figure*}
    \centering
    \includegraphics[width=1\linewidth]{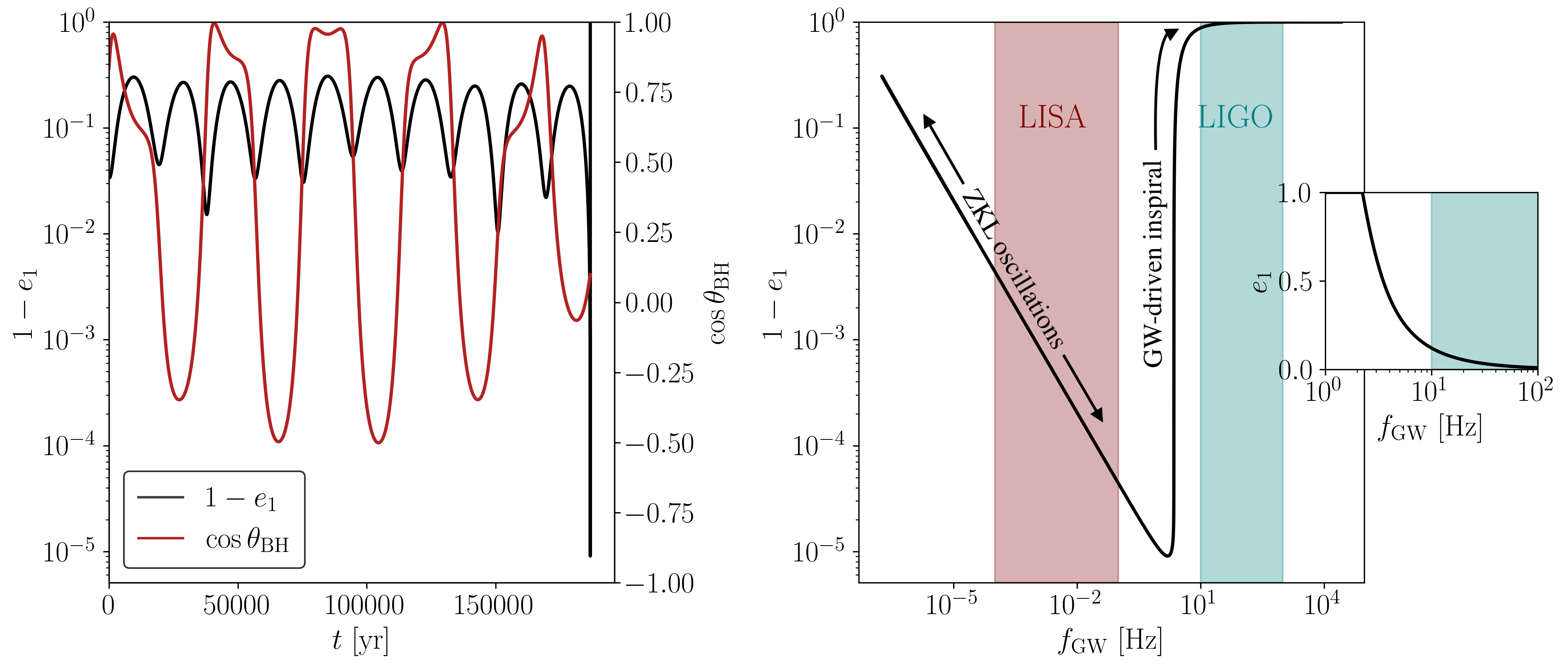}
    \caption{Evolution of an example tripe that leads to an NSBH merger with large BH spin-orbit misalignment and significant residual eccentricity. The left panel shows large-amplitude ZKL oscillations of the NSBH eccentricity $e_1$ (black line) as a function of time since NSBH formation. Meanwhile, the direction of the NSBH orbital angular momentum oscillates while the BH spin direction remains effectively constant, which leads to large changes of the spin-orbit angle (red line). The system merges after a few cycles at around $t\approx0.2\,\rm Myr$ with a spin-orbit misalignment close to $90\,\rm deg$. The right panel shows the eccentricity evolution as a function of the dominant GW frequency $f_{\rm GW}$. \new{Initial parameters of the system are $m_1 = 11.14\,\rm M_\odot$, $m_2 = 1.50\,\rm M_\odot$, $m_3 = 0.21\,\rm M_\odot$, $e_1 = 0.97$, $a_1 = 2.99\,\rm AU$, $e_2 = 0.19$, $a_2 = 27.57\,\rm AU$, $\mathbf{\hat e}_1 = ( 0.36, 0.93, -0.11 )^{\rm T}$, $\mathbf{\hat j}_1 = ( -0.53, 0.10, -0.84 )^{\rm T}$, $\mathbf{\hat e}_2 = ( -0.57, -0.39, 0.72 )^{\rm T}$, $\mathbf{\hat j}_2 = ( 0.68, 0.28, 0.68 )^{\rm T}$, and $\mathbf{\hat S}_{\rm BH} = ( -0.70, 0.52, -0.49 )^{\rm T}$.}}
    \label{fig:example}
\end{figure*}

    \item\label{item:non-adiabatic}\textit{Non-adiabaticity causes large spin-orbit misalignments:} The evolution of the BH spin can be characterised by the adiabaticity parameter $R=t_{\rm ZKL}/t_{\mathbf{S}_{\rm BH}}$ \citep{Anderson2017,Rodriguez2018} where $t_{\rm ZKL}$ is the ZKL timescale defined in Eq.~\eqref{eq:tLK} and
    \begin{equation}
        t_{\mathbf{S}_{\rm BH}}=\frac{\pi c^2a_1^{5/2}(1-e_1^2)}{2G^{3/2}\mu_{12}m_{12}^{1/2}}\left(1+\frac{3m_2}{4m_1}\right)^{-1}\label{eq:tS}
    \end{equation}
    is the de Sitter precession timescale of the BH spin vector about the inner binary angular momentum. Since all NSBH mergers are driven by a nearby tertiary companion ($a_2(1-e_2)/a_1\lesssim\mathcal{O}(10)$) and $R\propto \left(a_2\sqrt{1-e_2^2}\right)^3/a_1^4$ they occur in the non-adiabatic regime where $R\ll1$ (left panel of Fig.~\ref{fig:R_j}). In this regime, the BH spin evolves too slowly to effectively follow the direction of the inner orbital angular momentum vector through the relativistic spin-orbit coupling. Thus, the BH spin direction remains essentially fixed whereas the inner binary orbital angular momentum vector evolves rapidly due to the ZKL oscillations caused by the nearby tertiary companion \citep{Rodriguez2018}. As a result, the NSBHs can merge with a large spin-orbit misalignment of the BH, as exemplified by the system shown in Fig.~\ref{fig:example} where---simultaneously to the NSBH eccentricity---the direction of its orbital angular momentum undergoes large-amplitude oscillation and eventually forms a large angle with the BH spin at merger. Nevertheless, closely aligned spin-orbit orientations can occur even in the non-adiabatic regime. This is the case if the tertiary perturbation is so strong that NSBHs are driven to merge before a full ZKL cycle could be completed and the orbital angular momentum could substantially depart from alignment, which is shown in an example in Fig.~\ref{fig:example-2}.
    
    \new{Exemplified in Figs.~\ref{fig:example} and~\ref{fig:example-2} we also note that for most of the evolution the binaries oscillate at low GW frequency potentially being resolvable or contributing to a stochastic background in the LISA bandwidth \citep{Xuan2024,Knee2024,Xuan2025} until it reaches maximum eccentricity and undergoes an effectively isolated GW-driven inspiral.}

    \item\label{item:ecc}\textit{Tertiary-driven NSBH mergers can retain large residual eccentricity:} In general, tertiary-driven NSBH mergers occur if the ZKL effect induces a critically large eccentricity ($e_{1}\approx1)$ of the inner binary where the sudden energy-loss due to GW radiation at close periapsis shrinks the inner binary, decouples it from further perturbation from the tertiary, and is followed by an effectively isolated GW-driven inspiral \citep[e.g.,][]{Silsbee2017}. The critical eccentricity to undergo a GW-dominated inspiral can be either achieved before ZKL oscillations could be quenched by the relativistic apsidal precession of the inner binary or the latter becomes important before GW radiation dominates the binary evolution \citep{Rodriguez2018}. In the former case the inner binary only undergoes a few ZKL cycles before it ``plunges" directly into the highly eccentric GW-dominated regime, whereas in the latter case the binary is driven to a more ``gentle" merger caused by the cumulative effect of damped eccentricity excitations across many ZKL cycles. As a result, the binary can retain a large eccentricity upon merging in the former case and typically circularises below measurability in the latter case. Which case applies can be estimated by comparing the angular momentum $j_{\rm GW}$ where GW radiation dominates over the ZKL effect with the value $j_{\rm GR}$ where the latter is quenched by the Schwarzschild precession of the inner binary, which are given by \citep{Antonini2018,Rodriguez2018}
    \begin{align}
        j_{\rm GW}&=\left(\frac{170Gm_{\rm BH}m_{\rm NS}}{3c^5m_3}\frac{a_2^3j_2^3}{a_1}\nu^3\right)^{1/6},\label{eq:j_GW}\\
        j_{\rm GR}&=\frac{3G}{\pi c^2}\frac{(m_{\rm BH}+m_{\rm NS})^2}{m_3}\left(\frac{a_2j_2}{a_1}\right)^3\frac{1}{a_1}.\label{eq:j_GR}
    \end{align}
    For $j_{\rm GW}/j_{\rm GR}>1$ mergers occur without being affected by the Schwarzschild precession and can retain large residual eccentricities at merger. As shown in Fig.~\ref{fig:R_j}, this is the case for the majority of NSBH triples and large values $j_{\rm GW}/j_{\rm GR}$ correlate with a high probability to obtain large residual eccentricity $e_{20}>0.1$ evaluated at a dominant GW frequency of $f_{\rm GW}=20\,\rm Hz$. This is further illustrated in Fig.~\ref{fig:regimes} showing that most eccentric mergers are located near the diagonal dashed line referring to $j_{\rm GW}=j_{\rm GR}$. 

    \item\label{item:merger-times} \textit{Short delay times tie the tertiary-driven NSBH merger rate to the cosmic star formation:} The median delay time from NSBH formation to merger ranges from $1.95 \times 10^{4}$ (Bray) to $2.49 \times 10^{5}\,\rm yr$ (Zero) and the maximum 95th percentile is found to be $2.58 \times 10^{7}\,\rm yr$ (Zero). The short delay times can be understood by the large values of $j_{\rm GW}/j_{\rm GR}$ (see \ref{item:ecc}) which cause most mergers to occur in no more than a few octupole cycles $\sim\mathcal{O}(1)\times t_{\rm EKL}$ where $t_{\rm EKL}\sim t_{\rm ZKL}/\sqrt{\epsilon_{\rm oct}}$ is the timescale for octupole oscillations (often referred to as eccentric Kozai Lidov effect) \citep{Antognini2015}. In addition, massive stars evolve and form NSBHs after several Myr, which causes the total delay time from ZAMS to NSBH merger to be small on a cosmological scale and we expect the merger rate as a function of cosmological redshift to closely follow the cosmic star formation rate \citep{Madau2014}.

    \item\label{item:binary-interactions} \textit{Tertiary-driven NSBH mergers form from close interacting binary stars:} In order to survive the natal kick at NS formation the inner binaries need to be close. We find the inner semi-major axis distribution of surviving NSBH triples to peak at $a_1\sim \mathcal{O}(1)\,\rm AU$ and decline above a few $\sim10\,\rm AU$ (see Fig.~\ref{fig:SMA-histo}, red line). Thus, most of their progenitor stars underwent some sort of mass transfer (cf., blue patches) when they grew during stellar evolution and filled their Roche-lobe \citep{Eggleton1983}. Despite being driven by the gravitational perturbation from the tertiary companion, the properties of NSBH mergers (e.g., component masses and SN properties) are therefore strongly affected by the binary evolution of their progenitor stars \citep[cf.,][]{Marchant2024,Laplace2025}. This is different to the sample of NSBH mergers in triples obtained by \citet{FragioneLoeb2019a,FragioneLoeb2019b,Trani2022} who restricted their analyses to inner binaries with $a_1\gtrsim10\,\rm AU$ that effectively evolve as single stars. 
    Meanwhile, close and interacting inner binaries are more common \citep{Sana2012,Sana2014}, they are more likely to survive the NS natal kick, and the triple is less likely to be disrupted if the NS forms from a binary-stripped star, with reduced mass ejecta at SN explosion.
    Therefore, limiting the analyses to non-interacting triples overlooks most of the NSBH mergers.
    % Moreover, we argue below that the latter restriction leads to a much lower NSBH merger rate because the abundance of wide binaries is suppressed \citep{Sana2012} and they are easily disrupted by any non-zero natal kick.
\end{enumerate}
\begin{figure*}[ht!]
    \centering
    \includegraphics[width=1\linewidth]{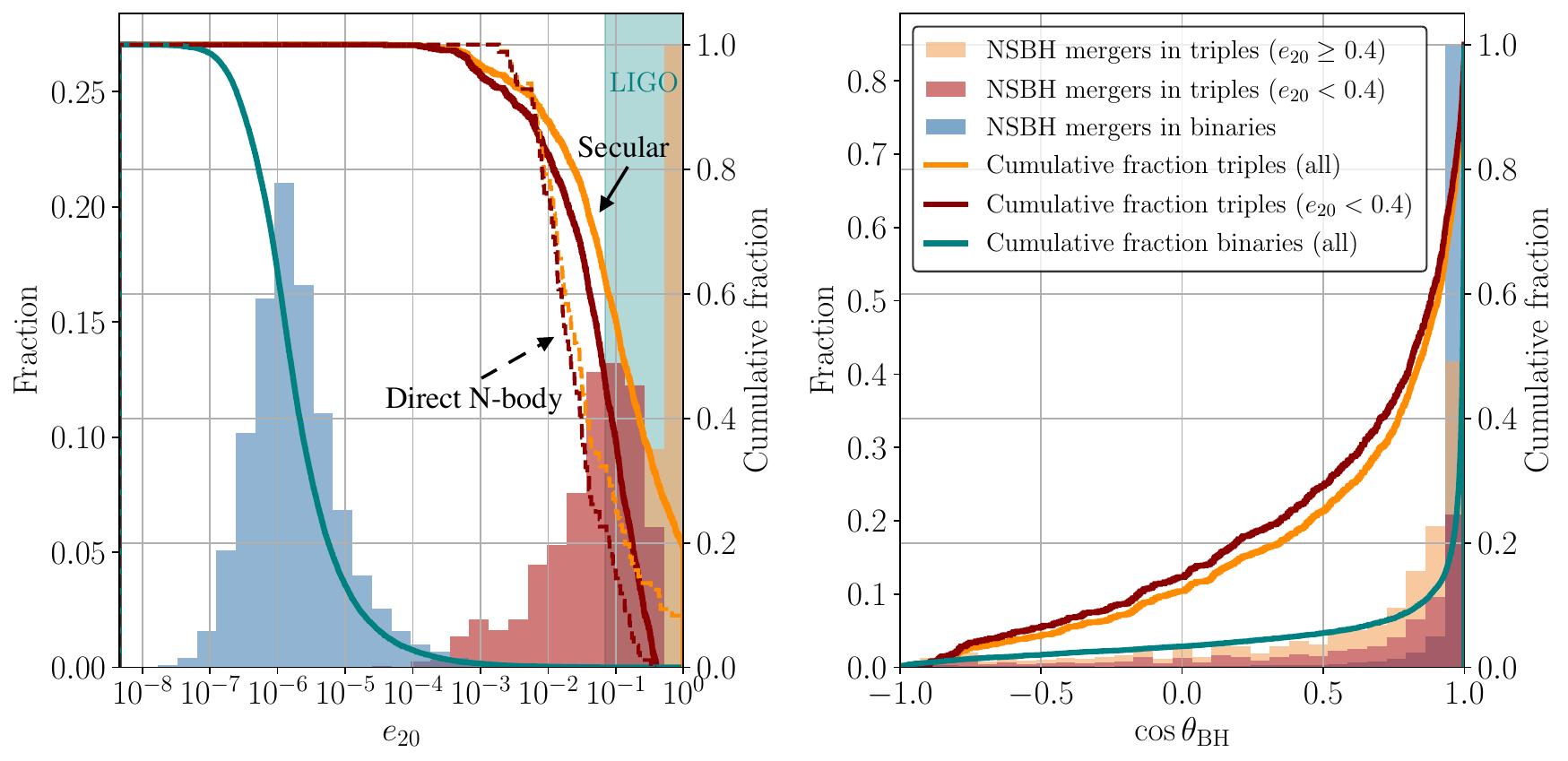}
    \caption{Distributions of the residual eccentricity at a dominant GW frequency $f_{\rm GW}=20\,\rm Hz$ \citep{Wen2003} of NSBH mergers (left) and the final BH spin-orbit misalignment (right). Red patches show NSBH mergers in triples below a threshold of $e_{20}<0.4$. Tertiary-driven NSBH mergers with higher eccentricities could be observed as burst-like sources and are shown by yellow patches. The distribution of isolated NSBH mergers without tertiary companions is shown by blue patches. Orange, red, and teal lines show the cumulative distributions of all NSBH mergers in triples, the subset of those with $e_{20}<0.4$, and isolated binary mergers, respectively. \new{Dashed lines show the cumulative eccentricity distribution from our direct $N$-body integration (see Sec.~\ref{sec:gravity}).} The teal shaded region on the left panel show the detectability limit $e_{20}\gtrsim0.07$ for LVK events.}
    \label{fig:distribution}
\end{figure*}

The properties \ref{item:oct}~--~\ref{item:binary-interactions} naturally result in a large fraction of tertiary-driven NSBH mergers with residual eccentricity and spin-orbit misalignment of the BH, as shown in Fig.~\ref{fig:distribution}\new{, in the Maxwellian\_100\_10 model}. The left panel shows that a high fraction ($\sim60\,\%$) tertiary-driven NSBH mergers retain a large residual eccentricity ($e_{20}\gtrsim0.1$) at $f_{\rm GW}=20\,\rm Hz$. The detectable range of eccentricities with the current network of ground-based GW detectors is about $0.07\lesssim e_{20}\lesssim0.4$, where below eccentricities are too small to be confidently distinguishable from circular orbits \citep[cf.,][on eccentric binary black hole mergers at $f_{\rm GW}=10\,\rm Hz$]{Lower2018,Nitz2020} and above (yellow patches) the GW signal becomes burst-like and is not well-modelled by some of current eccentric waveform approximants \citep[e.g.,][]{Nitz2020,planas2025}. Moreover, highly eccentric events risk being missed by the current searches for circular events by the LVK and would therefore not recognised by \citet{Morras2025} (although using approximants sensitive to higher eccentricities) who only reanalysed events which were identified by the LVK in the first place. Restricting the merger sample to the subset with $e_{20}\lesssim0.4$, we find a fraction of $\sim50\,\%$ with $e_{20}\gtrsim0.07$ (red line and patches). Obtaining an eccentricity at least as large as $e_{20}\ge0.145$ as inferred by \citet{Morras2025} for GW200105 has a probability of $\sim25\,\%$. Therefore, detecting such eccentric event out of up to \new{seven} NSBH mergers detected so far would be a very plausible outcome if triples are the dominant formation channel. Meanwhile, we expect a fraction $\sim96\,\%$ to have eccentricities $e_{20}\gtrsim10^{-3}$ which would be potentially detectable with the Einstein Telescope and Cosmic Explorer \citep{Saini2024}. In contrast, any event with detectable eccentricity would be wholly inconsistent with mergers originating from isolated NSBH binaries (blue patches and teal line) and would rule them out as a significant formation channel.  

\new{The above findings do not largely change when switching from the secular to the direct $N$-body integration, shown in dashed lines in the left panel of Fig.~\ref{fig:distribution}. We find that the fraction of mergers the fraction of eccentric mergers drops by a factor of $\sim2$ compared to the secular integration. Moreover, the total number of mergers is found to be smaller by a factor of $\sim5$. Individual systems often merge with one integration technique but not with the other, due to non-secular torques in the direct $N$-body simulation. However, the overall population statistics remain somewhat similar, although we do not find an increased total number of mergers and eccentric ones, as found in previous studies \citep[e.g.,][]{Antonini2014}.}

As shown on the right panel of Fig.~\ref{fig:distribution}, the distribution of the final BH spin-orbit angles of tertiary-driven NSBH mergers extends to large misalignment values ($\cos \theta_{\rm BH}\approx-1$) but peaks near alignment ($\cos \theta_{\rm BH}\approx+1$). This distribution can be understood by the two cases described in~\ref{item:non-adiabatic} and exemplified in Figs.~\ref{fig:example} and~\ref{fig:example-2}, where mergers with close spin-orbit alignment result from NSBH triples that could not have completed a full ZKL cycle before the mergers take place, whereas large misalignments are the result of large-amplitude oscillations of the orbital orientation across several ZKL cycles in the non-adiabatic regime. Obtaining retrograde spins with $\cos\theta_{\rm BH}<0$ as found for GW200115 \citep{Abbott2021} and GW230529 \citep{GW230529} is the case for $\sim14\,\%$ of the tertiary-driven mergers, which would be in agreement with $\sim1$~--~$2$ such events out of $\lesssim6$ detected NSBH candidates. In our models this is $4$~--~$5$ times more likely than for isolated binary mergers \citep[although see][]{Fragione2021}.

\begin{figure}
    \centering
    \includegraphics[width=1\linewidth]{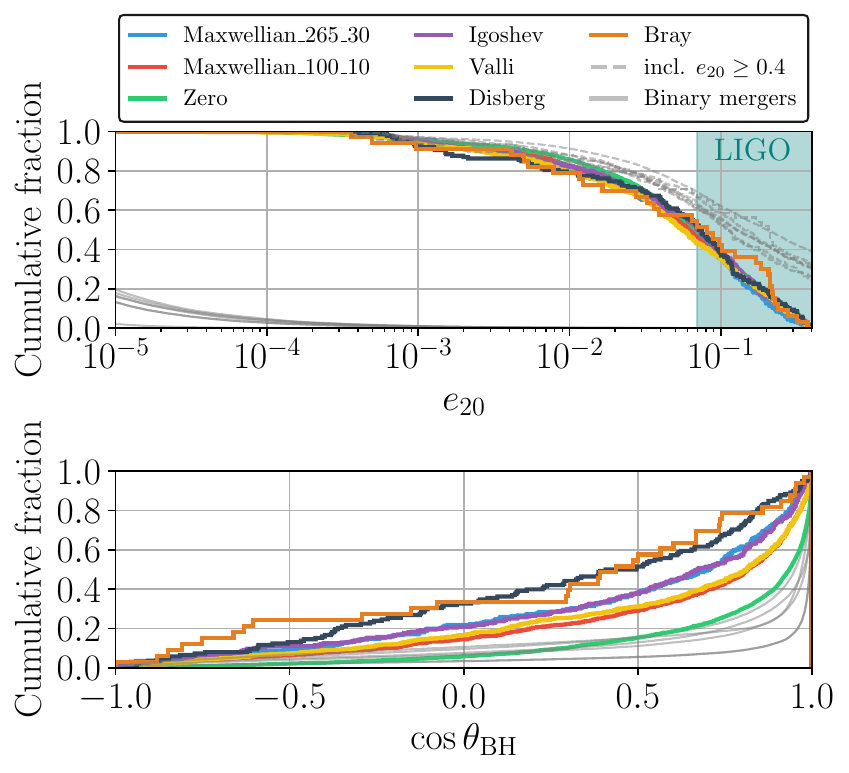}
    \caption{Cumulative distribution of the residual eccentricity at $f_{\rm GW}=20\,\rm Hz$ of tertiary-driven NSBH mergers (top panel) and of their spin-orbit misalignment (bottom panel). Different colours refer to our range of natal kick models, restricted to subset of mergers with $e_{20}<0.4$. Dashed grey lines include mergers above $e_{20}\ge0.4$, solid grey lines display mergers from isolated binaries.}
    \label{fig:models}
\end{figure}

Neither of our conclusions about the residual eccentricity and spin-orbit misalignment significantly change for the different kick models we explore. As shown in Fig.~\ref{fig:models}, the fraction of tertiary-driven NSBH mergers that retain significant eccentricity ($e_{20}>0.1$) remains largely unchanged (top panel), while the fraction of misaligned systems ($\cos\theta_{\rm BH}<0$) is only significantly lowered if zero kicks are assumed (lower panel). This reflects the fact that the properties of the surviving NSBH triple population is predominantly affected by the detrimental effect of mass-loss during NS formation (which remains unchanged across the different models) rather than the actual natal kick whose effect on the outer binary semi-major axis is damped by the low mass of the NS (see Eq.~\eqref{eq:a2})

Finally, the fact that large octupole effects (see \ref{item:oct}) enable mergers across a wide range of mutual inclinations (Fig.~\ref{fig:oct}) leads to a large fraction and rate of NSBH mergers which agrees with the current estimate \new{$\mathcal{R}_{\rm LVK}=9.1$~--~$84\,\rm Gpc^{-3}\,yr^{-1}$ from GW detections \citep{LIGO-O4-pop}}. We follow \citet{FragioneLoeb2019a} and calculate the NSBH merger rate by assuming a local star formation rate of $0.025\,\rm M_\odot\, Gpc^{-3}\,yr^{-1}$ so that the number of primary stars formed per unit mass is given by \citep{Bothwell}
\begin{equation}
    N(m_1)\,{\rm d} m_1 = 5.4\times10^6m_1^{-2.3}\,{\rm Gpc^{-3}\,yr^{-1}}{\rm d} m_1.
\end{equation}
Since our delay times are small (see~\ref{item:merger-times}) we can directly relate the local star formation rate to the local merger rate of NSBHs. Integrating across our investigated initial mass range $10\,{\rm M_\odot}<m_1<100\,{\rm M_\odot}$ we obtain 
\begin{equation}
    \mathcal{R}_{\rm NSBH}=2\times10^5f_3f_{\rm merge}\rm \, Gpc^{-3}\,yr^{-1},
\end{equation}
where we assume that a fraction $f_3=0.75$ of massive stars are in hierarchical triples on the ZAMS \citep{Moe2017} and determine the fraction $f_{\rm merge}=N_{\rm merge}/N_{\rm tot}$ from the number $N_{\rm merge}$ of tertiary-driven NSBH mergers formed out of $N_{\rm tot}$ stellar triples obtained in our simulations. Depending on the kick model we obtain
\begin{align}
\mathcal{R}_{\text{NSBH}} &=
\begin{cases}
5.46 & \text{Maxwellian\_265\_30} \\
22.87 & \text{Maxwellian\_100\_10} \\
219.24 & \text{Zero} \\
6.30 & \text{Igoshev} \\
22.81 & \text{Valli} \\
2.70 & \text{Disberg} \\
1.31 & \text{Bray} \\
\end{cases} \quad \text{Gpc}^{-3}~\text{yr}^{-1}\nonumber,
\end{align}
for the total rate of tertiary-driven NSBH mergers. Only including mergers below the threshold $e_{20}<0.4$ we obtain
\begin{align}
\mathcal{R}_{\text{NSBH}}^{e_{20} < 0.4} &=
\begin{cases}
4.02 & \text{Maxwellian\_265\_30} \\
15.91 & \text{Maxwellian\_100\_10} \\
133.67 & \text{Zero} \\
4.29 & \text{Igoshev} \\
15.88 & \text{Valli} \\
2.05 & \text{Disberg} \\
0.89 & \text{Bray} \\
\end{cases} \quad \text{Gpc}^{-3}~\text{yr}^{-1}\nonumber.
\end{align}

\new{The above range of NSBH merger rates illustrates the extent to which this channel is affected by natal kicks, but it should not be taken as the overall rate uncertainty of the triple scenario.  While natal kicks are a key factor, other poorly constrained processes also influence the survival of massive-star triples on their evolutionary path to forming NSBH triples. In particular, these include: (i) the stability of mass transfer in the inner binary, (ii) the stability of the triple is response to inner binary widening, (iii) and the criteria for the formation of a BH or a NS. In the model of \citet{Schurmann2024_stability} adopted here, the stability of mass transfer is directly related to the accretion efficiency ($\beta$): the more a star accretes, the more likely it is to rapidly expand in response and trigger instability. For our fiducial model, we adopted $\beta = 0.1$. motivated by the broad expectation for rotationally-limit accretion \citep{Langer2003}. Observations of post-interaction binaries suggest that a broader range of $\beta$ is possible \citep{Lechien2025}. We find that when instead assuming $\beta = 0.7$ ($0.0$), the number of NSBH triples in our population model decreases (increases) by a factor of $\sim4$. To gauge the effect of (ii), we explore a variation with a 'Jeans' mode for the ejection of the non-accreted matter, leading to more substantial inner orbit widening. We find that the number of NSBH triples decreases by only $\sim$10\%. Finally, to test the effect of (iii), we replace the prescription for BH/NS formation from \citet{Woosley2020} with that of \citet[][their Delayed variant]{Fryer2012} and \citet[][their Model A variant]{Maltsev2025}. We find that the predicted number of NSBH triples decreases by a factor $\sim2$ and $\sim3$, respectively. We therefore heuristically estimate that the NSBH rates given above for different natal kick variations are each subject to further uncertainty by a factor of $\sim10$.
Finally, our rates are likely subject to further uncertainty due to the poorly-constrained initial parameters of massive hierarchical triples. However, thanks in part to the significant octupole terms of the ZKL effect, our NSBH mergers are not restricted to any narrow parameter range of the triple orbits (e.g. no fine-tuning of inclinations). The most constrained parameter is the ratio $a_2/a_1$, ranging from $\sim$5 to $\sim$500 (Fig.~\ref{fig:SMA-histo}). While there are no strong constraints on the $a_2/a_1$ distribution for massive triples, among low-mass triples the $a_2/a_1$ was found to peak around a favourable value $\sim 10$ \citep{Shariat2025}.
}

In addition, some of the BHOB inner binaries may undergo CE evolution and form NSBH mergers essentially in isolation, regardless of the tertiary. Their contribution would increase the total rate and dilute somewhat the fraction of eccentric events. 
However, the question which systems survive CE evolution with post-CE orbits that are close enough to produce GW events is hugely uncertain. This is not only because of the general complexity of the CE phase \citep{Ivanova2020}, but also due to the poorly-understood interplay between the inner binary and the rich circumbinary matter ejected during the dynamical phase of CE, which may strongly influence the final orbit \citep{Ondratschek2022,Gagnier2023,Tuna2023,Wei2024, Vetter2024,Vetter2025}. The specific case of NSBHs might be particularly challenging for the CE channel if the dynamical in-spiral is followed by a slower mass exchange during the self-regulated phase \citep{Ivanova2011,Fragos2019,VignaGomez2022,Hirai2022}, which for unequal mass ratios of NSBHs is likely to widen the final orbit, easily to the point when the system no longer merges as a GW source \citep{Moreno2022,Wei2024}. The contribution of the CE channel to the NSBH merger population, if any, will be constrained by future GW events.
% with  the growing population of helium stars in binary systems with well-characterized orbits 

\section{Conclusion} 
\label{sec:summary}
Deciphering the origin of the compact object mergers observed through GW radiation is one of the major challenges in GW astronomy. The formation of NSBH mergers is particularly intriguing as the natal kick associated with the NS formation and their low mass ratio pose a challenge to proposed mechanisms, e.g., few-body scatterings in dense star clusters (see Sec.~\ref{sec:intro}). Meanwhile, formation from the evolution of isolated massive binary stars would be wholly inconsistent with the detection of eccentric NSBH mergers and is challenged by the detection of events with large BH spin-orbit misalignment, as recently inferred for GW200105 \citep{Fei2024,Morras2025,planas2025,Kacanja2025} and GW200115 and GW230529 \citep{Abbott2021,GW230529}, respectively.

Motivated by the fact that most massive stars reside in hierarchical triple systems \citep{Moe2017}, we investigated their evolution into NSBH mergers driven by gravitational perturbations from the tertiary companion via the ZKL mechanism. Among the various stages of stellar evolution, the second SN---which forms the NS---is the most disruptive event to the stability of the triple system. As a result, only compact NSBH triples tend to survive, characterised by properties \ref{item:oct}~--~\ref{item:binary-interactions}. Thus, we find the NSBH mergers in triples to occur at a rate $\mathcal{R}_{\rm NSBH}\simeq1$~--~$23\,\rm Gpc^{-3}\,yr^{-1}$ ($\mathcal{R}^{e_{20}<0.4}_{\rm NSBH}\simeq1$~--~$16\,\rm Gpc^{-3}\,yr^{-1}$)
% , where the range is refers to all non-zero kick models explored in this work,
and we obtain a considerable fraction of a few $10\,\%$ of residual eccentricity and large BH spin-orbit misalignment (Fig.~\ref{fig:distribution}). This is in excellent agreement with the rate of NSBH mergers inferred by the LVK as well as the number and the fraction of peculiar GW events showing tentative evidence for eccentricity and misalignment. This will be further constrained by NSBH detections in the upcoming GW observing runs. Further evidence for the triple channel may come from the growing sample of BHOB systems, possibly with tertiary companions (Klencki \& Stegmann, in prep.), that is set to substantially increase thanks to large-scale astrometric (GAIA), spectroscopic (e.g. LAMOST, SDSS), and microlensing surveys \citep[e.g., Vera Rubin,][]{Abrams2025,Shenar2024,ElBadry2024NewAR}.

If GW200105 is indeed eccentric or similar events are confidently detected in the future, our work suggests that hierarchical field triples are the dominant formation channel of NSBH mergers. Nonetheless, close binary star evolution on the inner orbit is a crucial aspect in determining the properties of the resulting NSBH mergers. On the one hand, a close separation ($a_1\lesssim\mathcal{O}(10)\,\rm AU$) of the inner binary is needed for a large number to survive the natal kick of the NS. Not including them explains why previous work by \citet{FragioneLoeb2019a} only found a high rate \new{$\mathcal{R}_{\rm NSBH}\simeq19\,{\rm Gpc}^{-3}\,{\rm yr}^{-1}$} of NSBH mergers from triples if zero kicks were assumed and \new{$\mathcal{R}_{\rm NSBH}\lesssim\mathcal{O}(10^{-2})\,{\rm Gpc}^{-3}\,{\rm yr}^{-1}$ if kicks taken into account. Meanwhile, previous work by \citet{Hamers2019} did find a high rate of NSBH mergers up to $\mathcal{R}_{\rm NSBH}\lesssim\mathcal{O}(10^{2})\,{\rm Gpc}^{-3}\,{\rm yr}^{-1}$, but since they did not fully account for three-body dynamics they could not report on the fraction of mergers with eccentricity or spin-orbit misalignment.} On the other hand, close binaries interact through mass transfer which critically affects the properties of the resulting compact remnants. We leave a detailed study investigating the implications of different assumptions about binary interaction for future work.

If binary black holes would predominantly merge in hierarchical triples too, we speculate it may provide a natural explanation for the seemingly smaller fraction of a few events out of about a hundred that have tentative evidence for eccentricity \citep{Shaw2022,Gupte2024,romeroshaw2025gw200208222617eccentricblackholebinary} and negative $\chi_{\rm eff}$ \citep{Abbott2023}. On the one hand, the mass-loss and natal kick associated with BH formation is generally expected to be much smaller than for NSs allowing much wider triple orbits to survive \citep[e.g.,][]{Antonini2017,Rodriguez2018}. On the other hand, the mass ratio distribution of binary black holes is generally closer to one compared to NSBHs, which would diminish the strength of octupole effects and narrow the parameter window for tertiary-driven mergers. As a result, we would expect the fraction of eccentric binary black hole mergers and with spin-orbit misalignment to be much lower than for NSBHs if both systems form in hierarchical triples.

%% Please use the acknowledgment and contribution environments. This will 
%% be anonomyized when the "anonymous" style option is used. 
\begin{acknowledgments}
We thank Patricia Schmidt, Fabio Antonini, Yossef Zenati, Alejandro Vigna-G\'omez, Ruggero Valli, Martyna Chruślińska, and Isobel Romero-Shaw for useful discussions and input. 
\end{acknowledgments}

\begin{contribution}
%%This section gives authors the space to recognize author contributions. The text inside this environment is NOT counted towards the total word quanta. At a minimum, manuscripts are expected to include this text:
Both authors contributed equally.
%% But authors are expected to provide more specific details, e.g. 
%%
%%SC was responsible for writing and sumathbfitting the manuscript.
%%WWM came up with the initial research concept and edited the manuscript.
%%OTS obtained the funding and edited the manuscript.
%%EBF provided the formal analysis and validation. He also edited the manuscript.
%%GEH Supervised the undergraduates, wrote the software and administers the project github and Zenodo repositories.
%%
%% Authors can use the Contributor Role Taxonomy (CRediT) at
%% for ideas on how write a good statement tailored to their needs.
\end{contribution}

\appendix

\section{Supplementary figures}

\begin{figure}
    \centering
    \includegraphics[width=.5\linewidth]{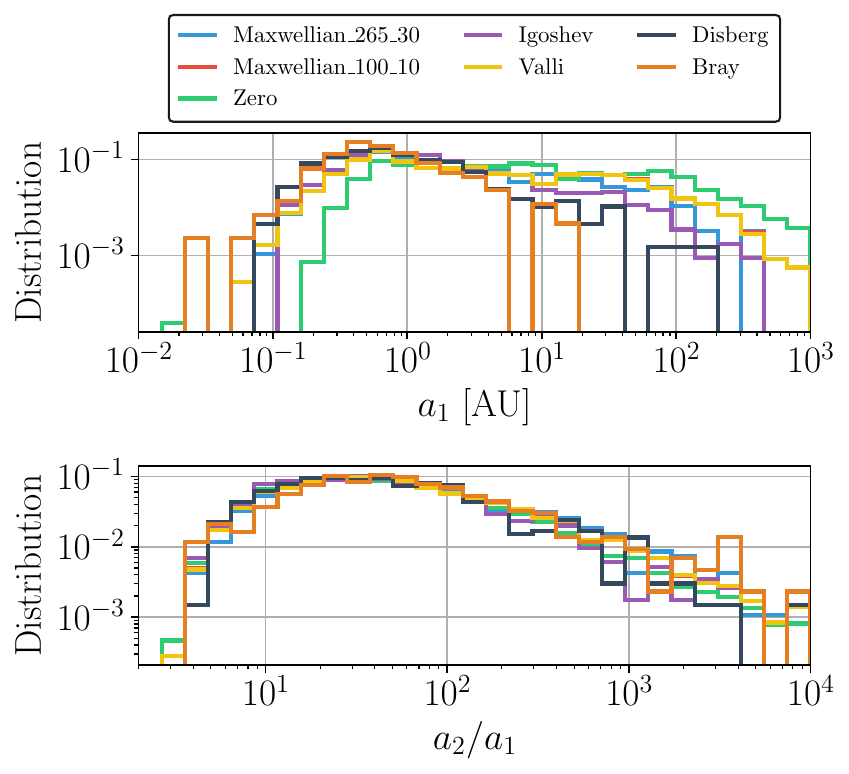}
    \caption{Distribution of inner semi-major axis $a_1$ (top panel) and semi-major axis ratio $a_2/a_1$ (bottom panel) before the second SN of systems that successfully form stable NSBH triples. Different colours refer to different natal kick models and the distributions are normalised with respect to~the total number of NSBH triples formed in each model.}
    \label{fig:a1_a1_models}
\end{figure}

\begin{figure*}
    \centering
    \includegraphics[width=1\linewidth]{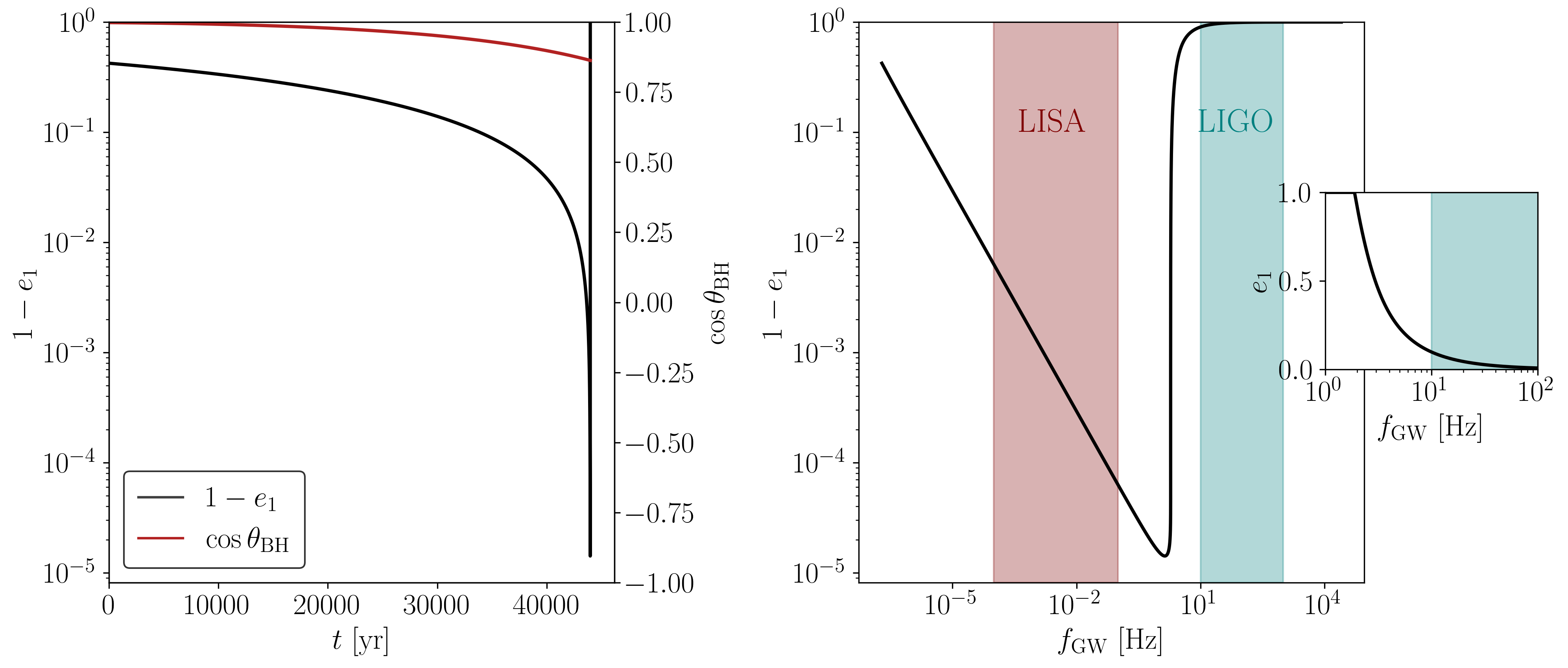}
    \caption{Same as Fig.~\ref{fig:example} for an example system that merges before fully completing a full ZKL cycle. As a result the inner orbital angular momentum could only slightly deviate from the BH spin direction. \new{The initial parameters are 
    $m_1 = 7.30\,\rm M_\odot$,
$m_2 = 1.50\,\rm M_\odot$,
$m_3 = 0.13\,\rm M_\odot$,
$e_1 = 0.58$,
$a_1 = 2.22\,\rm AU$,
$e_2 = 0.63$,
$a_2 = 58.22\,\rm AU$,
$\mathbf{\hat e}_1 = ( -0.92, 0.17, 0.35 )^{\rm T}$,
$\mathbf{\hat j}_1 = ( -0.33, 0.11, -0.94 )^{\rm T}$,
$\mathbf{\hat e}_2 = ( -0.51, 0.55, -0.66 )^{\rm T}$,
$\mathbf{\hat j}_2 = ( -0.19, 0.68, 0.71 )^{\rm T}$, and
$\mathbf{\hat S}_{\rm BH} = ( -0.33, 0.17, -0.93 )^{\rm T}$.}}
    \label{fig:example-2}
\end{figure*}

\bibliography{biblio}{}
\bibliographystyle{aasjournalv7}



\end{document}